\documentclass[a4paper,11pt,multicol]{article}
\pdfoutput=1
\usepackage{url}
\usepackage{algorithm}
\usepackage{cite}
\usepackage[noend]{algorithmic} 
\usepackage[utf8]{inputenc}
\usepackage{balance}
\usepackage[T1] {fontenc}
\usepackage{amsmath, amsthm}
\newtheorem{lemma}{Lemme}
 \newtheorem{theorem}{Théorème}
 \newtheorem{remark}{Remarque}
\usepackage[frenchb,english]{babel}%
\usepackage{color}
\newcommand{\daniel}[1]{\textcolor{red}{$<$}\scriptsize\textbf{#1}\textcolor{red}{$>$}\normalsize}
\newcommand{\danielsilenced}[1]{}
\newcommand{\kamel}[1]{\textcolor{blue}{$<$}\scriptsize\textbf{#1}\textcolor{blue}{$>$}\normalsize}
\newcommand{\owen}[1]{\textcolor{green}{$<$}\scriptsize\textbf{#1}\textcolor{green}{$>$}\normalsize}
\newcommand{\owensilenced}[1]{}
\newcommand{\cut}[1]{}
\newcommand{\kamelcut}[1]{}
\newcommand{\danielcut}[1]{}

\newcommand{\todo}[1]{\colorbox{yellow}{TODO}\small\textbf{#1}\normalsize}

\usepackage{subfig}

\usepackage{graphicx}

\graphicspath{{../BitmapIndexCpp/bash/}{.}{../BitmapIndexCpp/dbash/}{../BitmapIndexCpp/bash/sort_usincome/}{../BitmapIndexCpp/bash/sort_dbgen/}{../BitmapIndexCpp/bash/sort_netflix/}{../BitmapIndexCpp/bash/sort_dblp/}{../BitmapIndexCpp/bash/block_usincome/}{../BitmapIndexCpp/bash/block_dbgen/}{../BitmapIndexCpp/bash/block_dblp/}{../BitmapIndexCpp/bash/block_netflix/}{../BitmapIndexCpp/synthetic/}}

\usepackage{ifpdf}
\ifpdf
\else

\fi

\usepackage{ulem}

\usepackage{url}
\let\urlorig\url
\renewcommand{\url}[1]{%
  \begin{otherlanguage}{english}\urlorig{#1}\end{otherlanguage}%
}

\NoAutoSpaceBeforeFDP

\usepackage[cyr]{aeguill}

\date{}

\begin{document}

\selectlanguage{frenchb}

\title{Tri  de la table de faits et  compression des index bitmaps avec alignement sur les mots}

\author{\centering Kamel Aouiche$^1$, Daniel Lemire$^1$ et Owen Kaser$^2$  
}

\maketitle

\begin{abstract}

Les index bitmaps sont souvent utilisés pour indexer des données multidimensionnelles. Ils utilisent
principalement l'accès séquentiel aux données, tant à l'écriture qu'à la lecture.
Les bitmaps peuvent être compressés pour réduire le coût des entrées/sorties tout en minimisant l'utilisation du microprocesseur.
%
Les techniques de compression les plus efficaces sont fondées sur la compression par plage (\textit{run-length encoding})
telle que la compression alignée sur les mots. 
Ce mode de compression permet d'effectuer rapidement
des opérations logiques (AND, OR) sur les bitmaps. Cependant, la compression par plage
dépend de l'ordre des faits. 
Nous proposons donc d'exploiter le tri de la table de faits afin d'améliorer
l'efficacité des index bitmaps. Le tri lexicographique et par code de Gray, ainsi que le tri par bloc, sont  évalués. 
Selon nos résultats expérimentaux, un simple tri lexicographique peut produire un index
mieux compressé (parfois deux fois plus petit) et qui peut être plusieurs fois plus rapide.
Le temps requis par le tri apporte un surcoût au temps total d'indexation, mais ce surcoût
est amorti par le fait que l'indexation d'une table triée est plus rapide.
L'ordre des colonnes peut avoir une influence déterminante sur l'efficacité du tri lexicographique~:
il est généralement préférable de placer les colonnes ayant plus de valeurs distinctes au début.
Le tri par bloc sans fusion est beaucoup moins efficace qu'un tri complet. 
De plus, le tri par code de Gray n'est pas supérieur au tri lexicographique
dans le cas de la compression alignée sur les mots.






\end{abstract}
\hspace{1cm}\textbf{Mots-clefs}~: index bitmap, optimisation, compression, codes de Gray
\vspace*{0.5cm}
\begin{quote}
\textbf{Fact Table Sorting and Word-Aligned Compression for Bitmap Indexes}
\small

Bitmap indexes are frequently used to index multidimensional data. They rely mostly on sequential input/output.
Bitmaps can be compressed to reduce input/output costs and minimize CPU usage.
The most efficient compression techniques are based on run-length encoding (RLE), such
as Word-Aligned Hybrid (WAH) compression.
This type of compression accelerates logical operations (AND, OR) over the bitmaps.
However, run-length encoding is sensitive to the order of the facts.
Thus, we propose to sort the fact tables. 
We review lexicographic, Gray-code, and block-wise sorting.
We found that a lexicographic sort improves compression---sometimes generating indexes
twice as small---and make indexes several times faster. While sorting takes time,
this is partially offset\owensilenced{stupid irregular verb} by the fact that it is faster to index a sorted table.
Column order is significant: it is generally preferable to put the columns having
more distinct values at the beginning. A block-wise sort is much less efficient than
a full sort. Moreover, we found that Gray-code sorting is not better than 
lexicographic sorting when using word-aligned compression.
\end{quote}
\vspace*{0.5cm}
{\small
kamel.aouiche@gmail.com, lemire@acm.org, o.kaser@computer.org\\
$^1$ LICEF, Universit\'e du Qu\'ebec \`a Montr\'eal, 100, rue Sherbrooke Ouest,\\ Montr\'eal (Qu\'ebec), H2X 3P2 Canada\\
$^2$  University of New Brunswick, 100 route Tucker Park,\\ Saint Jean, Nouveau-Brunswick E2L 4L5 Canada
}

\twocolumn

\section{Introduction}
Les entrepôts de données stockent de grands volumes de données~:  il n'est pas rare
de trouver des entrepôts faisant plus de 100\,To~\cite{monashbigtera2007} et parfois même 
jusqu'à 10\,Po~\cite{xldb07---fr}.
On trouve des tables faisant plus d'un billion de faits ($10^{12}$)~\cite{xldb07---fr}.
La construction rapide d'index multidimensionnels sur de grands volumes est donc indispensable pour accélérer
 l'accès aux données. La technique des index bitmaps (ou index binaires) est l'une des solutions les plus communément utilisées
 avec l'arbre B et les
tables de hachage. 

Sans compression, certains index bitmaps
sont peu pratiques. Par exemple, l'indexation d'un seul attribut comportant 10~millions valeurs
et 100~000~valeurs distinctes nécessiterait un téraoctet.
Dans le but de réduire la taille des index bitmaps et de les rendre plus efficaces, 
plusieurs méthodes de compression ont été proposées. Nous citons la compression par plage~\cite{133210},
 la compression alignée sur les octets BBC (\textit{Byte-Aligned Bitmap Compression})~\cite{874730}, et
 la compression Lempel-Ziv (LZ77)~\cite{chan1998bid}. 
Une autre technique de compression, plus compétitive que la simple compression par plage, est la compression
alignée sur les mots WAH (\textit{Word-Aligned Hybrid})~\cite{wu2006obi}. En comparaison
avec LZ77 et BBC, WAH fournit des index parfois beaucoup plus rapides~\cite{502689}. 

\begin{figure} 
\centering 
\includegraphics[width=0.95\columnwidth]{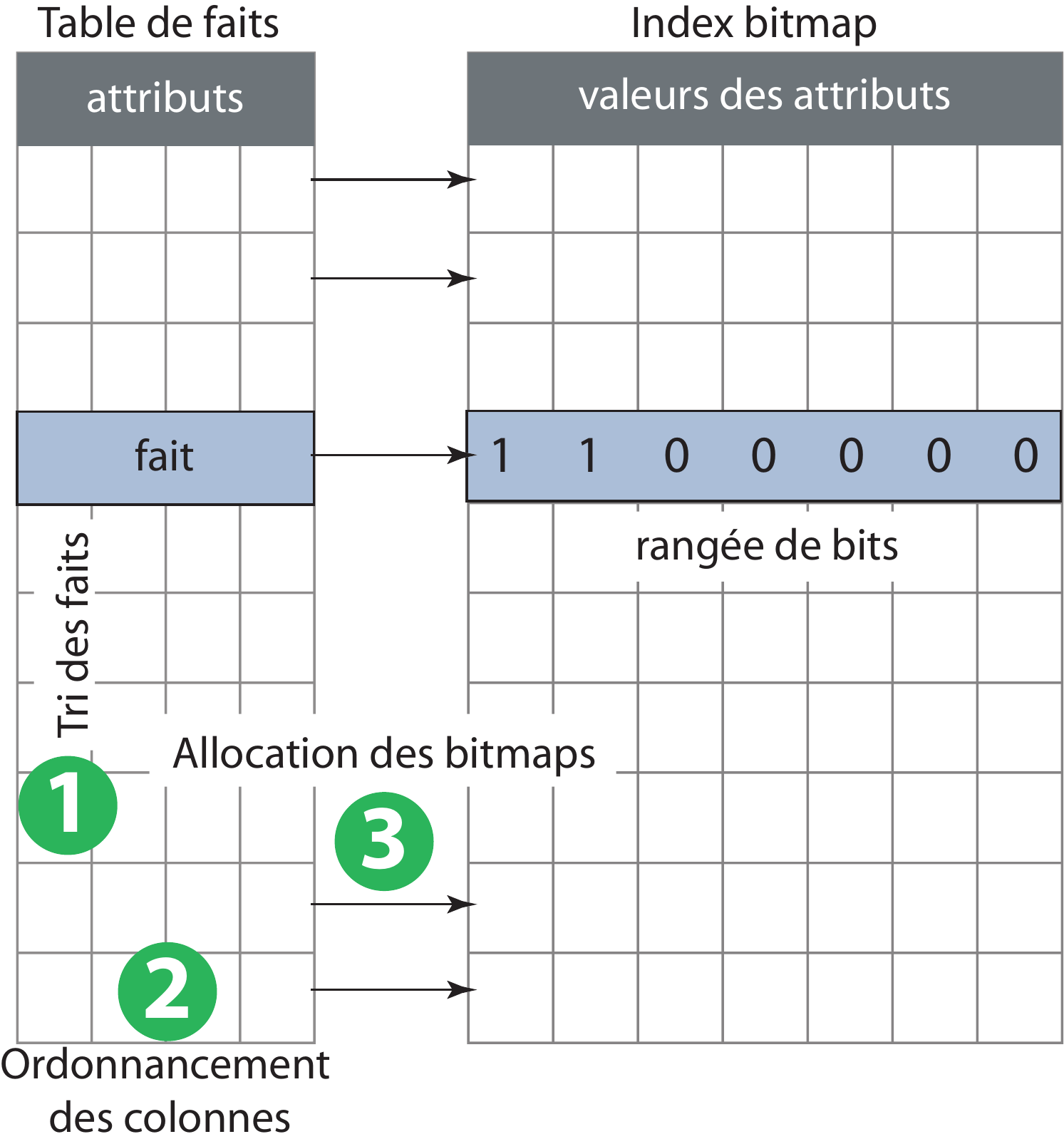}
\caption{\label{fig:sort} Application du tri et de l'ordonnancement des dimensions pour améliorer les performances des index bitmaps.}
\end{figure}

Nous étudions dans ce papier 
plusieurs méthodes de tri pour améliorer la compression 
et le temps de calcul des index bitmaps. Le tri, comme le montre la Figure~\ref{fig:sort}, peut 
être appliqué sur les faits 
ou sur les codes binaires assignés aux faits. 
Nous nous intéressons plus particulièrement au tri
lexicographique et au tri par code de Gray. 
De plus, nous discutons de l'ordonnancement des dimensions selon leurs cardinalités
afin d'améliorer le coût de stockage et de calcul des index bitmaps (voir la Figure~\ref{fig:sort}).
L'étude réalisée dans ce papier porte particulièrement sur les tables de faits. 
Cependant, elle valable dans un contexte de table autre qu'une table de faits.   
À notre connaissance, il s'agit de la première étude portant sur cette question.
Jusqu'à présent, seul le tri des faits par code de Gray a été 
proposé~\cite{pinar05,pinarunpublished---fr}.
Cependant, ces derniers travaux n'ont considéré que le cas d'index de petites tailles~: le plus 
volumineux des index comporte 2~millions de faits et 900~bitmaps. La description des auteurs du 
tri par code de Gray suppose en effet que l'index bitmap soit d'abord matérialisé
sans compression ; ce qui est coûteux en espace de stockage et  temps de calcul. 
Dans notre cas, nous considérons des dimensions ayant plus de 50\,000~bitmaps et 
des tables ayant plus de 100~millions de faits.  
Nous étudions l'effet que le tri préalable des faits a sur les temps
de création des index et  d'interrogation des données.

\section{Index bitmaps}

\subsection{Principes de base}

Les entrepôts de données représentent souvent leurs données sous la forme d'une table 
de faits (Tableau~\ref{tab:faits}) et des tables de dimensions (Tableau~\ref{tab:dimensions}). 
Une table de faits est composée de plusieurs colonnes (attributs : clés étrangères et mesures) 
et de plusieurs rangées (faits). 
Une dimension donnée peut prendre plusieurs valeurs~: la dimension «~Ville~» 
peut prendre les valeurs «~Montréal~», «~Paris~», etc. 
Les tables de faits peuvent atteindre des volumes importants, allant 
de millions de faits dans le cas de petites entreprises, à des milliards de faits ou encore 
plus dans le cas de grandes entreprises. 
Afin de supporter efficacement les requêtes en ligne sur de telles tables, 
des index sont souvent nécessaires.

\begin{table}
\centering
\caption{\label{tab:examples}Une table de faits, des tables de dimensions et un index bitmap}

\subfloat[Table de faits\label{tab:faits}] {
\begin{tabular}{cccc}
id-Ville & id-Véhicule & id-Couleur & quantite \\ \hline
1 & 1 & 1 & 2 \\
2 & 2 & 1 & 1\\
3 & 1 & 2 & 10 \\
4  & 1 & 1 & 3\\
5 & 1 & 3 & 1 \\
1 & 2 & 1 & 2 \\
6 & 2 & 1 & 1 \\
\end{tabular}
}
\hfill{}
\subfloat[Tables de dimensions\label{tab:dimensions}] {
\begin{tabular}{cc|cc|cc}
id & Ville & id & Véhicule & id & Couleur \\ \hline
1 & Montréal & 1 & Voiture & 1 & Bleue \\
2 & Paris & 2 & Autobus & 2 & Rouge \\
3 & Londres &  &  & 3 & Verte \\
4 & Québec  & &  & & \\
5 & Tokyo & & & & \\
6 & Lyon & & & & \\
\end{tabular}
}
\hfill{}
\subfloat[Index bitmap simple\label{tab:basiquebitmap}] {
\begin{tabular}{ccc}
100000 & 10 & 100 \\
010000 & 01 & 100 \\
001000 & 10 & 010 \\
000100 & 10 & 100 \\
000010 & 10 & 001 \\
100000 & 01 & 100 \\
000001 & 01 & 100 \\
\end{tabular}
}
\end{table}

Le premier exemple d'un index bitmap dans un moteur de base de données (MODEL~204) a été commercialisé
pour l'IBM~370 en 1972~\cite{658338}.
On trouve maintenant les index bitmaps dans de nombreux moteurs de base
de données, dont les bases de données Oracle. Le principal avantage de ce mode d'indexation
est de favoriser l'accès séquentiel aux disques tant 
à l'écriture
qu'à la lecture des données~\cite{jurgens2001tbi}.

Le principe de base d'un index bitmap est de représenter les faits de telle manière
à ce que les requêtes se traduisent par de simples opérations logiques (AND, OR). 
Pour chaque valeur $v$ d'un attribut $a$ d'une dimension, un index bitmap est composé d'un tableau de
booléens (ou bitmap) ayant autant de bits que de faits et dont les positions des bits mis à 1 correspondent 
aux faits pouvant être joints avec cet attribut.  
Le Tableau~\ref{tab:basiquebitmap} montre un exemple d'un index bitmap. Par exemple, le tableau de booléens
1000010 correspond au prédicat «~Ville~=~Montréal~» 
; qui n'est
vrai que pour le premier et le sixième faits.  
De la même manière, le prédicat  «~Véhicule~=~Autobus~» donne le tableau de booléens 0100011.
À titre d'illustration, les résultats de la requête  
«~Véhicule~=~Autobus AND Ville~=~Montréal~» sont calculés en réalisant le AND logique sur les deux
tableaux de booléens précédents.


Les index bitmaps sont rapides parce qu'au lieu de lire toutes les valeurs
possibles d'un attribut donné, 
on peut ne lire qu'aussi peu
qu'un bit par fait et par attribut, ou même moins si l'on exploite des
techniques de compression. Par ailleurs, un microprocesseur  peut
calculer 32 ou 64 opérations logiques sur un bitmap en une seule instruction. Le regroupement
 des bits d'un bitmap par paquets de 32 ou de 64 bits ne peut donc qu'améliorer grandement l'efficacité des index bitmaps.

Alors qu'on considère parfois que les index bitmaps sont surtout appropriés pour
les attributs de petite cardinalité, comme le sexe ou le statut marital, Wu et al. 
ont montré qu'ils sont également efficaces lorsque les cardinalités des données
sont très élevées~\cite{oraclevivekbitmap---fr,wu2006obi}. Par ailleurs,
les index bitmap permettent d'indexer plusieurs dimensions sans mal, alors que
les performances des index multidimensionnels en arbre tels que les arbres R, se
dégradent rapidement lorsque le nombre de dimensions augmente~\cite{671192}.

\subsection{Encodage des index bitmaps}

Il est possible de réduire le nombre de tableaux de booléens, donc de bitmaps, lors de l'indexation de données de forte cardinalité. Un index bitmap simple d'un seul attribut ne comporte qu'une seule valeur vraie par fait («~Montréal~» devient 100000) et
un bitmap par valeur distincte de l'attribut.
Observons  qu'étant donnés $L$~bitmaps, il existe ${L \choose 2} = L (L -1 )/2$~paires de bitmaps. En représentant les valeurs de la colonne non plus par la position de la valeur vraie, mais plutôt par la position de deux valeurs vraies («~Montréal~» devient 1100), on peut utiliser beaucoup moins de bitmaps (voir le Tableau~\ref{tab:examples1ofk})~: 
$(\sqrt{8N+1}+1)/2$~bitmaps suffisent pour représenter $N$~valeurs distinctes. Par exemple, seuls 2~000~bitmaps sont nécessaires pour représenter 2~millions de valeurs distinctes.

De manière plus générale, l'encodage $k$-of-$N$ permet de n'utiliser que $L$~bitmaps pour représenter ${L \choose k}$~valeurs distinctes. En contrepartie, au lieu de charger un seul bitmap correspondant  à une valeur d'attribut impliquée dans une requête, il faut désormais en charger $k$ bits, et effectuer une opération logique AND sur l'ensemble des données.
  
  Lorsqu'il existe une hiérarchie sur une dimension, on peut utiliser une variante de l'encodage $k$-of-$N$
  qui soit particulièrement appropriée~: on utilise les premiers
  $h_1$~bitmaps pour le premier niveau de la hiérarchie, les $h_2$~bitmaps suivants pour la deuxième
  et ainsi de suite. Par exemple, si les 3 premiers bitmaps sont dédiés au premier niveau de la hiérarchie
  (Canada, France, Belgique), et que les 5~bitmaps suivants sont utilisés pour distinguer l'une des 5~villes
  dans chacun des pays, «~Paris~» pourrait avoir le code 10010000 alors que «~Montréal~»
  pourrait avoir le code 01010000. Il s'agit d'une forme de dénormalisation~\cite{kimball1996dwt}
  qui permet d'éviter à avoir un index distinct sur les données après un roll-up sur une dimension.
 
 On peut aussi regrouper les valeurs des attributs dans des lots~\cite{354819,1155030,stockinger2004esb} lorsque
  leur nombre est très important. Par exemple, au lieu d'utiliser un bitmap pour chaque rue de Paris,
  on peut utiliser un bitmap par quartier. Il existe des encodages pour supporter les requêtes
   par plage lorsque la cardinalité des attributs est importante~\cite{chan1999ebe}.
 
Lorsqu'on réduit le nombre de bitmaps, on augmente aussi la densité de l'index, c'est-à-dire le ratio des valeurs
 vraies par rapport aux valeurs fausses. Si l'encodage 2-of-$N$ génère toujours un index bitmap non compressé 
 beaucoup plus petit que l'encodage 1-of-$N$, la même chose n'est pas nécessairement vraie si l'on compresse l'index.
 
 Lorsqu'on souhaite indexer une table \cut{de faits}%
  à l'aide  de l'encodage $k$-of-$N$ pour $k>1$, il arrive qu'on puisse
 générer un index particulièrement inefficace sur les colonnes ayant peu de valeurs distinctes. Par exemple,
 si seulement 5~valeurs distinctes sont observées, un encodage simple (1-of-5) n'utilisera que 5~bitmaps
 alors qu'un encodage 2-of-$N$ exigera 4~bitmaps. Il est certain qu'un encodage 2-of-4 sera moins efficace
 qu'un encodage 1-of-5. Nous appliquons donc l'heuristique suivante pour nos tests. Toute colonne ayant 5~valeurs
 distinctes ou moins est limitée à l'encodage 1-of-$N$. Toute colonne ayant 21~valeurs distinctes ou moins
 est limitée aux encodages 1-of-$N$ et 2-of-$N$. Toute colonne ayant 85~valeurs distinctes ou moins
 est limitée aux encodages 1-of-$N$, 2-of-$N$ et 3-of-$N$. 

\begin{table}
\centering
\caption{\label{tab:examples1ofk}Comparaison entre l'encodage 1-of-$N$ et l'encodage 2-of-$N$}
\subfloat[1-of-$N$]{
\begin{tabular}{ccc}
100000 \\
010000  \\
001000  \\
000100 \\
000010 \\
100000 \\
000001  \\
\end{tabular}
}
\subfloat[2-of-$N$]{
\begin{tabular}{ccc}
1100 \\
1010\\
1001 \\
0110\\
0101\\
1100 \\
0011 \\
\end{tabular}
}
\end{table}

\subsection{Compression des index bitmaps}

Le principe de la compression par plage est le codage des longues plages
de valeurs identiques par un nombre indiquant le nombre de répétitions suivi de 
la valeur répétée. Par exemple, la séquence 11110000 devient 4140. 
Comme il est fréquent que les bitmaps comprennent un grand nombre de zéros
ou de uns sur de longues plages, la compression par plage s'avère particulièrement appropriée. 
Cependant, il existe une autre raison pour laquelle
cette compression est efficace : elle rend plus rapide l'exécution des opérations logiques.
Considérons  deux bitmaps $B_1$ et $B_2$ comportant chacun $\vert B_1 \vert$
et  $\vert B_2 \vert$ valeurs vraies. Des algorithmes simples, comme 
l'Algorithme~\ref{algo:genrunlengthand}, permettent
de calculer les opérations $ B_1 \land B_2$ et  $ B_1 \lor B_2$
en temps $O(\vert B_1 \vert + \vert B_2 \vert)$.
En utilisant des variantes de cet algorithme, nous obtenons le lemme suivant.

\begin{algorithm}[tb]
\small
\begin{algorithmic}
\STATE \textbf{INPUT:} deux bitmaps $B_1$ et $B_2$ 
\STATE $i \leftarrow$ itérateur sur les valeurs vraies de $B_1$
\STATE $j \leftarrow$ itérateur sur les valeurs vraies de $B_2$
\STATE $S$ ensemble représentant les valeurs vraies de $B_1 \land B_2$ (initialement vide)
\WHILE{$i$ ou $j$ n'est pas arrivé à la fin du bitmap}
\STATE $ a \leftarrow $ position($i$) $<$ position($j$)
\STATE $ b \leftarrow $ position($i$) $>$ position($j$)
\WHILE {position($i$) $\neq$ position($j$) }
\IF{(position($i$) $>$ position($j$)) }
\STATE incrémenter l'itérateur $i$ si possible, sinon sortir de la boucle
\ELSE
\STATE incrémenter l'itérateur $j$ si possible, sinon sortir de la boucle
\ENDIF
\ENDWHILE
\IF{position($i$) $=$ position($j$)}
\STATE ajouter position($i$) à $S$
\ENDIF
\ENDWHILE
\end{algorithmic}
\caption{\label{algo:genrunlengthand}Calcul du AND logique de deux bitmaps
}
\end{algorithm}

\begin{lemma}\label{rlelogical} Étant donnés deux bitmaps $B_1$ et $B_2$ compressés par plage et comportant
$|B_1|$ et $|B_2|$ valeurs vraies, on peut calculer les opérations logiques  
$B_1 \land B_2 $, $B_1 \lor B_2 $, $B_1 \oplus B_2 $,  $B_1 \land not(B_2) $,
$B_1 \oplus not(B_2) $ en un temps $O(\vert B_1 \vert + \vert B_2 \vert)$. 
\end{lemma}

Observons que le nombre de séquences de valeurs vraies doit être égal (plus ou moins un)
au nombre de séquences de valeurs fausses.

L'inconvénient de l'Algorithme~\ref{algo:genrunlengthand} est que les opérations logiques 
se font bit-par-bit alors que le microprocesseur opère sur des mots.
Au lieu de manipuler chaque bit un à un, il peut-être plus judicieux de manipuler des mots. 
Nous distinguons deux types de mots : les mots propres ne comportant que des bits vrais (1x11)
ou des bits faux (0x00), et les autres, appelés mots impropres. Nous pouvons
coder toute séquence de mots propres de même nature par un entier suivi d'un bit pour distinguer
les deux types de mots propres (1x11 et 0x00), alors que les séquences de mots impropres
peuvent être codées par un entier déterminant la longueur de la séquence, suivi par 
les mots impropres représentés \textit{in extenso}.

Wu et al.~\cite{wu2006obi} appliquent cette idée sur l'encodage WAH, mais 
plutôt sur des mots de 31~bits.
Ils représentent les séquences de mots propres du même type
par un mot dont le premier bit est faux, le deuxième bit indique le type
du mot propre (0x00 ou 1x11) et les 30~bits suivants sont utilisés pour stocker
la longueur de la séquence. Les mots impropres de 31~bits sont stockés \textit{in extenso},  tout en
mettant le premier bit à la valeur vraie pour indiquer qu'il s'agit d'un mot impropre.
 Dans le pire des cas, l'encodage WAH peut donc générer des bitmaps
dont la taille est supérieure au bitmap non-compressé par un facteur de 32/31 ce qui
représente une expansion de plus de 3\,\%.
Notre méthode de compression, EWAH (\textit{Enhanced Word-Aligned Hybrid}), utilise
un encodage avec des mots de 32~bits\footnote{Par souci
 de simplicité, nous ne présentons notre méthode que
pour les mots de 32~bits.}. Les mots impropres sont stockés
\textit{in extenso} alors que les mots propres sont compressés à l'aide d'un  mot-marqueur. Le 
mot-marqueur comprend trois informations~: un bit est utilisé pour donner le type
de mots propres, 16~bits pour stocker le nombre de mots propres du type
donné, et les 15~bits suivants pour compter le nombre de mots impropres
suivant la séquence de mots propres. L'inconvénient de notre  implémentation
est qu'il faut toujours débuter un bitmap par un mot-marqueur ; que le bitmap débute
par une séquence de mots propres ou pas~\footnote{Cette contrainte est purement technique.}. Dans le cas extrême où nous avons de très longues séquences de mots
propres, le codage WAH peut s'avérer plus efficace parce qu'il utilise un mot
tous les $2^{30}$~mots propres de 31~bits alors que nous utilisons un mot 
tous les $2^{16}$~mots propres de 32~bits~: ce scénario se produirait  si nous avions 
plusieurs fois $32 \times 2^{16}$
~faits consécutifs sans qu'une valeur
d'un attribut donnée n'est observée
\footnote{Pour une table \cut{de faits}%
 comportant 100~millions faits
et indexée par 20~000 bitmaps, ce phénomène pourrait ajouter, dans le pire des cas, 7\,Mo à la taille
de l'index encodé par EWAH par rapport à WAH. Comme un tel index fait plus de 180\,Mo dans nos tests, l'effet
est donc de moins de 4\%.}.
 Par ailleurs, dans la mesure où la table \cut{de faits}%
est suffisamment grande, il n'est pas possible pour l'encodage EWAH de générer des bitmaps
compressés ayant une taille supérieure au bitmap non compressé (à 0,1\% près), contrairement à l'encodage WAH.

Avec la compression alignée sur les mots, pour effectuer une opération 
logique comme AND, il suffit de lire à la fois les mots propres de type 1x11 et impropres.
Le lemme suivant est démontré de manière semblable au Lemme~\ref{rlelogical}.

\begin{lemma}Étant donnés deux bitmaps $B_1$ et $B_2$ compressés avec alignement sur les mots
 et comportant
$|B_1|$ et $|B_2|$ mots non nuls, incluant 1x11 et tout mot impropre,  on peut alors calculer les opérations logiques  
$B_1 \land B_2 $, $B_1 \lor B_2 $, $B_1 \oplus B_2 $,  $B_1 \land not(B_2) $,
$B_1 \oplus not(B_2) $ en un temps $O(\vert B_1 \vert + \vert B_2 \vert)$. 
\end{lemma}

L'espace occupé par un bitmap $B$ compressé par alignement sur les mots 
est dans $O(|B|)$ et  $O(|B|)= O(\textrm{vrai}(B))$ où $\textrm{vrai}(B)$ est le nombre de valeurs
vraies dans un index bitmap.  
De plus, lorsque les index bitmaps sont suffisamment creux, c'est-à-dire qu'ils comportent beaucoup
plus de valeurs fausses que de valeurs vraies, la plupart des mots propres n'a que des bits
faux (0x00) : le coût des opérations est dans $O(\textrm{impropre}(B_1)+ \textrm{impropre}(B_2))$.

\section{Tri de la table de faits pour améliorer la compression des index bitmaps}

\subsection{Formalisation du problème}

Si on suppose que l'on exploite la compression par plage, on peut aisément formaliser
le problème d'ordonnancement des rangées (ensemble des bits assignés à un fait) dans un index bitmap. Si
$d(r,s)$ est le nombre de bits qui diffère entre les rangées 
$r$ et $s$, il faut
trouver l'ordre des rangées $r_i$ qui minimise  $\sum_i d(r_i,r_{i+1})$.
Par exemple, si deux faits sont représentés par les rangées de bits 0110 et 0011, 
alors la différence est de 2. 
Pinar et al.~\cite[Théorème 1]{pinar05} ont montré qu'on peut réduire ce problème
de l'ordonnancement des faits au problème du voyageur, qui est lui-même NP-difficile. Ce qui
montre que le problème de l'ordonnancement n'est pas plus difficile que le problème du voyageur.
Nous constatons que comme $d$ satisfait l'inégalité 
du triangle, il est possible de calculer une approximation 1,5-optimale en un temps cubique~\cite{Christofides1976}
en utilisant des approximations au problème du voyageur.
Les auteurs affirment par ailleurs que le problème
est NP-difficile. L'objet du paragraphe qui suit est de démontrer que c'est effectivement le cas.

\begin{theorem}\label{thm:trivialTCS}Le problème de l'ordonnancement des
rangées dans un index bitmap compressé par plage
est NP-difficile.
\end{theorem}
\begin{proof}
Le problème du chemin Hamiltonien consiste 
à trouver un chemin 
 qui visite chaque sommet exactement une fois. Ce problème est NP-complet. Inspirés par Johnson et al.~\cite{johnson2004clb},
nous réduisons le problème du chemin Hamiltonien au problème d'ordonnancement
des rangées. Considérons un graphe quelconque  $G$. 
Construisons une matrice  $M$ comme suit. Chaque colonne de la
matrice représente un sommet et chaque rangée 
un arc. Pour chaque rangée, indiquons par la valeur 1 les colonnes correspondant aux
sommets joints par un arc donné. Toutes les autres composantes
de la rangée sont nulles. Lorsque deux arcs $r$ et $s$ ont un sommet en commun 
$d(r,s)=1$, sinon $d(r,s)=2$. En minimisant   $\sum_i d(r_i,r_{i+1})$,
nous trouvons un chemin Hamiltonien s'il existe.
\end{proof}

Cependant,  si le nombre de bitmaps $L$ est petit et que seul
le nombre de faits est grand, on peut démontrer
qu'une solution en temps polynomial est possible : c'est certainement
vrai s'il n'y a qu'un seul bitmap.

Si on généralise l'ordonnancement des rangées au cas de la compression alignée sur les
mots, le problème n'est pas toujours NP-difficile, même lorsque le nombre de bitmaps
est grand. Par exemple, si la taille des mots
est égale au nombre de rangées, 
le problème devient trivial.  Formellement, on peut généraliser le problème 
ainsi~: il faut ordonner les rangées dans un index bitmap en tenant compte
que le stockage d'une séquence de mots propres  de même type (0x00 ou 1x11)
coûte $w$~bits, alors que le stockage  de tout autre mot coûte $w$~bits.
Lorsque $w$ est une petite constante, le problème est NP-difficile.

\begin{theorem}
Le problème de l'ordonnancement des rangées dans un index bitmap compressé par alignement des mots est NP-difficile si le nombre de bits par mot ($w$) est une constante.
\end{theorem}
\begin{proof}
Considérons le cas où chaque rangée du bitmap est répétée un multiple de $w$. 
Il est donc possible d'ordonner les rangées de telle manière à ce que nous n'ayons que des mots propres (1x11 et 0x00). 
Il existe une solution optimale de ce type de problème. Dans ce cas, le problème est équivalent à un ordonnancement des mots propres de des deux types afin de réduire leur compression par plage, un problème NP-difficile selon le Théorème~\ref{thm:trivialTCS}.
\end{proof}

\subsection{Méthodes de tri}
\label{sec:methodedetri}
L'ordre  lexicographique est défini sur des séquences  
d'éléments~: une séquence $a_1,a_2, \ldots$ est plus petite qu'une autre
séquence $ b_1,b_2,\ldots $ si et seulement
s'il existe $j$ tel que $a_j< b_j$ et $a_i=b_i$ pour $i<j$.
La commande Unix «~sort~» permet de trier des fichiers
plats en utilisant le tri lexicographique. Cette opération est relativement
peu coûteuse. Par exemple, sur l'ordinateur utilisé pour nos tests 
(voir la Section~\ref{sec:implementation}), un fichier plat
comportant 5~millions de lignes et plus de 120\,Mo peut être trié en 
moins de 10\,secondes. L'instruction SQL «~ORDER BY~» permet aussi 
de trier les faits de manière lexicographique en spécifiant
plusieurs colonnes~: SELECT * ORDER BY t1, t2, t3.

Pour montrer que le tri de gros fichiers n'apporte pas de surcharge élevée, 
considérons le problème où nous avons
suffisamment de mémoire interne pour stocker $M$~éléments 
et un total de $n$~éléments ($ n> M$) à trier.
Le tri en mémoire externe, tel mis en \oe{}uvre par la commande  «~sort~»,
procède en deux étapes~\cite{yiannis2007ctf}~: 

\begin{enumerate}
\item  $\lceil n/M \rceil$~blocs
de $M$~éléments sont lus, triés et 
écrits sur le disque un à un.
\item Une partie de la mémoire ($x$~éléments) est allouée pour l'écriture sur
le disque, alors que le reste de la mémoire est utilisé pour lire les premiers $z=\lfloor (M-x)/\lceil n/M \rceil \rfloor$~éléments
de chaque blocs. Les éléments lus sont stockés dans un tas. Le plus petit élément du tas est
retiré et déposé dans le tampon mémoire d'écriture. Cette dernière opération est répétée ; dès qu'un des $\lceil n/M \rceil$~tampons de lecture est vide, il est rempli à nouveau par
les $z$ prochains éléments du bloc. Dès que le tampon d'écriture est plein, il est écrit sur le disque.
\end{enumerate}
Ainsi, en théorie, avec une capacité de mémoire de $M$~éléments, on peut trier près de $M^2$~éléments en deux
passages, sans accès aléatoire au disque. Un ordinateur disposant de quelques
gigaoctets de mémoire devrait être à même de trier un ou deux téraoctets de données de
cette manière. Si la mémoire disponible est insuffisante pour faire
deux passages, des variantes de cet algorithme 
nécessitent $\lceil \log_{M} n \rceil$~passages.

Il peut être judicieux de remplacer le tri  d'une très grande table, 
par un tri par bloc sans fusion des blocs triés. On évite ainsi des lectures
non-séquentielles sur le disque. Un tri par bloc peut être calculé plus 
rapidement et est facilement parallélisable.


Dans notre implémentation des index bitmaps, lorsque nous assignons des bitmaps
aux valeurs, nous le faisons par défaut en préservant l'ordre alphabétique des valeurs des attributs.
Ainsi, dans un encodage 2-of-$N$, si le code 101000 est assigné à la valeur «~arbre~»,
le code 110000 peut être assigné à la valeur «~aaron~» 
(voir l'Algorithme~\ref{algo:allocalpha}). De cette manière, le tri lexicographique
de la table\cut{ de faits}, et le tri des rangées de bits dans l'index bitmap, sont
deux opérations équivalentes.

\begin{algorithm}
\begin{algorithmic}
\STATE \textbf{INPUT}: Les valeurs d'un attribut en ordre alphabétique
\STATE \textbf{INPUT}: $k$ le nombre de bitmaps alloué pour chaque valeur
\STATE \textbf{INPUT}: $N$ le nombre de bitmaps
\STATE $a \leftarrow \{1,2,\ldots,k\}$
\FOR{chaque valeur $v$ de l'attribut}
\STATE alloue les bitmaps $a_1, a_2,\ldots, a_k$ à $v$
\STATE $i \leftarrow k$
\WHILE{$a_i = L-(k-i)$ et $i > 1$}
\STATE $i \leftarrow i - 1$
\ENDWHILE
\STATE $a_i \leftarrow a_i +1, a_{i+1} \leftarrow a_i +2,\ldots  $
\ENDFOR
\end{algorithmic}
\caption{\label{algo:allocalpha}Allocation alphabétique des bitmaps aux valeurs d'attributs}
\end{algorithm}

Le tri par code de Gray~\cite{pinar05}  est défini sur
des tableaux de bits. En traitant les séquences de bits comme des codes
de Gray, la séquence $a_1,a_2, \ldots$ est inférieure à la séquence 
$b_1,b_2,\ldots $ si et seulement
s'il existe $j$ tel que~:
\begin{align*} 
a_j = \textbf{impair}(a_1, a_2, \ldots, a_{j-1}),
\end{align*}
$b_j \not = a_j$, $a_i=b_i$ pour $i<j$, où 
$\textbf{impair}(a_1, a_2,$ $ \ldots, a_{j-1})$
vaut 1 s'il y a un nombre impair de valeurs unitaires dans la liste 
$a_1, a_2, \ldots, a_{j-1}$ et 0 sinon. Le Tableau~\ref{tab:tri}
présente un exemple. 

\begin{remark}Étant donné un ensemble contenant toutes les rangées de bits
possibles sont représentées, un tri par code de Gray fait en sorte
que la distance de Hamming entre deux rangées qui se suivent soit au plus égale à un.
En d'autres termes, le tri par code de Gray fournit un ordonnancement optimal
des rangées dans le cas où  les codes de Gray possibles sont présents.
\end{remark}

Dans le cas où la table \cut{de faits}%
 ne comporte qu'une seule colonne et qu'on utilise
l'encodage 1-of-$N$, le tri par code de Gray est alors équivalent au tri
lexicographique. En effet, si   $a_i=b_i$ pour $i<j$ et $a_j \not = b_j$,
on a alors $a_i=b_i=0$ pour $i<j$ ; d'où 
$\textbf{impair}(a_1, a_2, \ldots, a_{j-1})=0$. De plus,
$a_1,a_2, \ldots < b_1,b_2,\ldots $ si et seulement
s'il existe $j$ tel que $a_j = 0$,
$b_j =1$, et $a_i=b_i=0$ pour $i<j$.

S'il y a plusieurs colonnes et que l'on utilise l'encodage 1-of-$N$ pour
chaque colonne, le tri par code de Gray est alors équivalent à un
tri lexicographique où l'ordre alphabétique est inversé une colonne sur
deux. Ainsi, si le fait «~Montréal, Voiture~» précède le fait
«~Paris, Voiture~», le fait «~Montréal, Voiture~» précède aussi le fait
«~Montréal, Autobus~». Il est plus difficile de faire la même 
analyse lorsque l'encodage $k$-of-$N$ pour $k>1$ est utilisé. Cependant,
si on utilise l'encodage $k_i$-of-$N$ pour la colonne $i$,
il n'en reste pas moins que l'ordre du tri par code de Gray est
inversé pour la colonne $c$ lorsque $\sum_{j < c} k_j$ est impaire
par rapport à l'ordre qu'il aurait eu si  $\sum_{j < c} k_j$ était paire.


Pour éviter d'avoir à faire le moindre tri par code de Gray sur les rangées
tout en bénéficiant des propriétés des codes de Gray,
on peut modifier l'allocation des bitmaps. Si la colonne $i$ suit l'encodage
$k$-of-$N$, on peut trier par code de Gray les codes $k$-of-$N$ 
et allouer le premier code à la première valeur de l'attribut (en ordre alphabétique)
et ainsi de suite. Ainsi, si on utilise un encodage 2-of-4, on allouera les
codes dans l'ordre suivant 0011, 0110, 1100, 1010, et 1001. Il suffit ensuite de trier
la table \cut{de faits}%
 de manière lexicographique. Cette approche n'est applicable
que lorsque $k>1$.  

Afin de trier les rangées de bits  par code de Gray, sans matérialiser
un index bitmap non compressé, nous avons utilisé
un arbre~B sur disque~\cite{qdbm---fr}. L'arbre~B a pour clés les positions des bits dont la valeur est vraie, et 
pour valeur de ces clés la ligne du texte correspondant à un fait dans le fichier plat qui décrit la table de faits.
Par exemple, si un fait représenté par les bits 100010001 dans l'index bitmap,
nous utiliserons les entiers 0, 4, et 8 pour clés. En utilisant un
encodage $k$-of-$N$ et des entiers de 32~bits, une table de faits ayant
$d$~dimensions aura donc des clés de $4kd$~octets. 
En comparaison, 
 une table de faits nécessitant
100~000~bitmaps aura des clés utilisant 12\,Ko si on stockait les rangées
de bits sans compression (par exemple,  100010001). Afin de réduire davantage la taille
de l'arbre~B, nous avons compressé les n\oe{}uds de l'arbre par LZ77. 
L'arbre~B 
traite
 les séquences de bits comme des codes de Gray. Après avoir construit
l'arbre, il suffit alors de traverser ses valeurs de manière croissante. 
Le résultat est une table de faits triée.
Sur de grosses tables, cette approche est au moins 100~fois plus lente que le tri
lexicographique. C'est en partie à cause du processus 
coûteux inhérent à la construction des arbres~B. Dans nos tests, nous 
n'appliquerons pas le tri par code de Gray sur nos gros fichiers de données.


\begin{table}
\centering
\caption{\label{tab:tri}Comparaison entre les deux méthodes de tri}
\subfloat[lexicographique]{
\begin{tabular}{ccc}
0 & 1 & 1\\
0 & 1 & 1\\
1 & 0 & 1\\
1 & 0 & 1\\
1 & 1 & 0\\
1 & 1 & 0\\
1 & 1 & 1\\
1 & 1 & 1\\
1 & 1 & 1\\
\end{tabular}
}
\subfloat[codes de Gray]{
\begin{tabular}{ccc}
0 & 1 & 1\\
0 & 1 & 1\\
1 & 1 & 0\\
1 & 1 & 0\\
1 & 1 & 1\\
1 & 1 & 1\\
1 & 1 & 1\\
1 & 0 & 1\\
1 & 0 & 1\\
\end{tabular}
}
\end{table}

\section{Mise en \oe{}uvre et résultats expérimentaux}
\label{sec:implementation}
Pour des fins d'expérimentation, nous avons développé un moteur d'index bitmap en C++. Notre
moteur lit un fichier plat dont les champs sont séparés par une virgule ou un autre caractère
et génère l'index bitmap correspondant.
Il est capable d'indexer des fichiers comprenant des millions de valeurs distinctes d'attributs
et des milliards de lignes (faits).
Pour assurer la reproduction de nos résultats, nous rendons disponible notre 
logiciel\footnote{\url{http://code.google.com/p/lemurbitmapindex/}.}.

L'index bitmap est stocké dans un seul fichier. L'entête du fichier comprend les méta-données
permettant d'associer les valeurs d'attribut aux bitmaps. Nous ne considérons pas le temps
du chargement en mémoire de cet entête lors de la présentation la vitesse des requêtes,
et de sa taille dans le calcul de la taille des index.
L'indexation se fait en deux passages. Un premier passage sur le fichier
de données permet de calculer un histogramme. Nous ne prenons pas en compte le temps de cette opération
lorsqu'on présente le temps de création des index.
L'histogramme est matérialisé sur disque afin d'éviter ce passage lorsqu'on 
réindexe le fichier avec des paramètres différents. 
Le deuxième passage écrit dans
le fichier d'index des blocs de 256\,Mo~: l'index est donc partitionné
horizontalement. Le processus d'écriture du fichier-index est toujours séquentiel, 
sans aucun accès aléatoire. Chaque bloc comporte un entête donnant la position 
en octet au sein du bloc de chaque bitmap~: l'entête a donc une taille de $4L$~octets
où $L$ désigne le nombre de bitmaps. Le reste du bloc est constitué des bitmaps compressés
écrits l'un à la suite de l'autre. Lors des requêtes, seuls les bitmaps requis sont lus.
Nous fixons le nombre de bits par mot à 32.

Pour une indexation rapide lorsque le nombre de bitmaps est élevé,
on évite d'accéder à chaque bitmap pour chaque rangée du fichier de données ;
ce qui donnerait une complexité $O(nL)$.
L'Algorithme~\ref{algo:owengenbitmap} est exécuté sur chaque partition
horizontale et fonctionne en temps $O(nkd +L)$ où $k$ est l'encodage choisi, 
$n$ est le nombre de faits dans la partition,
et $d$ le nombre de dimensions ou de colonnes. Nous utilisons le fait qu'on
puisse ajouter plusieurs mots propres de même type à un bitmap compressé par alignement de mots en 
un temps constant.

\begin{algorithm}[tb]
\small
\begin{algorithmic}
\STATE $B_1,\ldots, B_L$, $L$ bitmaps compressés 
\STATE $B_i.t \leftarrow B_i.b \leftarrow \omega_i \leftarrow 0$ pour $1 \leq i \leq L$.
\STATE $c\leftarrow 1$ \COMMENT{ compteur de rangées}
\STATE $\mathcal{N} \leftarrow \emptyset$ \COMMENT{$\mathcal{N}$ champs}
\FOR{chaque rangée de la table}
\FOR{chaque attribut de la rangée}
\FOR {chaque bitmap  $i$ correspondant à la valeur d'attribut}
\STATE met à vrai le $(c \bmod w)^{\textrm{e}}$~bit du mot $\omega_i$
\STATE $\mathcal{N} \leftarrow \mathcal{N} \cup \{i\}$
\ENDFOR
\ENDFOR
\IF{$c$ est un multiple de $w$}
\FOR{$i$ in $\mathcal{N}$}
\STATE ajoute $c/w-\vert B_i \vert  - 1$~mots propres (0x00) à $B_i$
\STATE ajoute le mot $\omega_i$ au bitmap $B_i$
\STATE $\omega_i \leftarrow 0$
\ENDFOR
\STATE $\mathcal{N} \leftarrow \emptyset$
\ENDIF
\STATE  $c\leftarrow c+1$
\ENDFOR
\FOR{$i$ in \{1,2,\ldots,L\}}
\STATE $c/w-\vert B_i \vert  - 1$~mots propres (0x00) à $B_i$ 
\ENDFOR
\end{algorithmic}
\caption{\label{algo:owengenbitmap}Algorithme de construction des bitmaps employé. Par souci
de simplicité, on suppose que le nombre de rangées est un multiple du nombre de bits dans un mot.}
\end{algorithm}

Nous avons utilisé pour les tests le compilateur GNU GCC version 4.0.2 sur une machine Apple Mac~Pro, 
dotée de deux processeurs Intel Xeon double c\oe urs tournant à une 
cadence de 2,66\,GHz et disposant de 2\,GiB de mémoire vive. 
Les temps d'exécution rapportés 
correspondent au temps écoulé et comprennent donc tout le temps
de calcul et le temps des entrées-sorties.

\subsection{Jeux de données}
\label{sec:dim}

Nous utilisons quatre ensembles de données~: Census-Income~\cite{KDDRepository---fr}, DBGEN~\cite{DBGEN---fr}, DBLP~\cite{DBLPXML---fr}, et Netflix~\cite{netflixprize---fr}. Le fichier Census-Income est composé de 43~dimensions et  de près de 200~mille faits. DBGEN est constitué de données synthétiques alors que les trois autres ensembles sont des données réelles.
La table de faits retenue pour DBGEN comprend 12~millions de faits et de 16~dimensions. 
La table de faits DBLP a été créée à partir du fichier XML représentant les centaines de milliers d'articles
et de communications répertoriés par le site DBLP\footnote{\url{http://dblp.uni-trier.de/}}. Le fichier XML est converti en un fichier CSV où chaque fait est une paire auteur-article. 
Les dimensions sont le type (compte-rendu, revue, etc.), le nom de la revue ou de la conférence, le nom de l'auteur, le titre et l'année de publication. Les cardinalités des dimensions sont 6, 3~574, 395~578, 528~344 et 49, respectivement. Il y a environ 1,5~millions de faits. 
La table de faits de Netflix comprend 4~dimensions MovieID, UserID, Rating, Date dont les cardinalités sont respectivement 17~770, 480~189, 5, et 2~182 et un total de plus de 100~millions de faits. Avant le traitement des données Netflix, qui sont regroupées en 17~700 petits fichiers (un fichier par film), nous avons créé un fichier plat unique. L'ordre initial de tous les fichiers de données est rendu aléatoire avant chaque test à l'aide de la commande Unix du type «~cat -n myfile.csv | sort --random-sort | cut -f 2-~». 
Il arrive que les jeux de données soient préalablement triés et pour rendre compte de l'apport du tri, il faut effectuer une permutation aléatoire avant chaque test. À cause de cette permutation, certains résultats peuvent varier légèrement 
d'une expérience à l'autre. 
Le Tableau~\ref{tab:caractDataSet} résume les principales caractéristiques de nos fichiers de données.

\begin{table}
    \centering
    \begin{scriptsize}
    \begin{tabular}{c|rrr|} \cline{2-4}
     & \# de faits & cardinalité & taille\\ \hline
    \multicolumn{1}{|c|}{\textbf{Census-Income}} & 199~523 & 43 & 99.1\,Mo   \\ 
    \multicolumn{1}{|c|}{\textbf{DBGEN}} & 13~977~981  & 16 & 1.5\,Go \\ 
    \multicolumn{1}{|c|}{\textbf{DBLP}} & 1~372~259 & 5 &  156\,Mo\\
    \multicolumn{1}{|c|}{\textbf{Netflix}} & 100~480~507  & 4 &   2.6\,Go \\ \hline
    \end{tabular}
    \end{scriptsize}
    \caption{Caractéristiques des jeux de données utilisés}\label{tab:caractDataSet}
\end{table}

Pour nos tests, nous avons sélectionné trois dimensions pour chaque ensemble de données en prenant
soin de choisir des dimensions ayant des cardinalités différentes.
Par exemple, les dimensions utilisées pour Census-Income sont : \textit{age} (d1), 
\textit{education} (d2), et  
\textit{migration code-change in msa} (25$^{\textrm{e}}$~champ) (d3).
Le Tableau~ \ref{tab:carUSIncome} donne la cardinalité de chacune de ces dimensions. Par convention,
la dimension ayant la plus faible cardinalité est toujours d1. 
Census-Income et DBLP ont tous les deux une dimension d3 dont la taille avoisine 
le nombre de faits (voir Tableaux~\ref{tab:caractDataSet} et~\ref{tab:carUSIncome}).
Netflix a de loin les plus petites dimensions en comparaison avec le nom de faits.

\begin{table}[!h]
    \centering
    \begin{scriptsize}
    \begin{tabular}{|c|r|r|r|r|} \cline{2-5}
    \multicolumn{1}{c|}{}   & \textbf{Census-Income} & \textbf{DBGEN} & \textbf{DBLP} & \textbf{Netflix} \\ \hline
    d1 & 91 & 7 & 49 & 5 \\ 
    d2 &  1~240 & 11 & 3~574 & 2~182\\ 
    d3 & 99~800 &  400~000 & 528~344 & 17~770\\ \hline
    \end{tabular}
    \end{scriptsize}
    \caption{Cardinalités des dimensions choisies pour chaque jeu de données}\label{tab:carUSIncome}
\end{table}

Pour d'autres expérimentations, nous avons utilisé des données synthétiques ayant
une distribution uniforme ou de Zipf et comportant environ 5~000~faits. 
Les données Zipfiennes sont composées de dimensions ($d \in \{1,\ldots,6\}$). Le paramètre de Zipf $s$, qui contrôle le 
biais des données, est dans  $\{0.5, 1.0, 1.5, 2.0\}$.
Les données de distribution uniforme ont au maximum 12 dimensions.
Chaque valeur d'attribut est un entier choisi d'une manière uniforme et aléatoire. 
La première dimension peut prendre ses valeurs parmi 100 valeurs distinctes, 
la deuxième parmi $100r$ et la troisième parmi $100r^2$, 
et ainsi de suite, où $r$ ne peut prendre que les valeurs 1 ou 2 et 
ne sert qu'à contrôler la variété dans le nombre des valeurs distinctes. 
Certains de ces jeux de distribution uniforme contiennent à la fois des attributs indépendants et dépendants.


Les attributs {}\og dépendants\fg{} sont obtenus de la manière suivante. 

Soit une rangée dont les valeurs des attributs $a_1, a_2, \ldots, a_d$ sont indépendantes. Une valeur dépendante 
est calculée par $\sum a_i p_i$, où  la valeur de $p_i$ est égale à 1 avec une probabilité de 0,2 et à 0 sinon. 
Si $\sum p_i = 0$, nous choisissons uniformément une valeur dans  $\{1,\ldots,100\}$.

Après avoir généré les données de cette manière, nous permutons aléatoirement les colonnes. 
Cela permet de s'assurer que les dimensions  de cardinalité élevée ou petite ne se 
retrouvent pas en tête des colonnes et que les attributs 
dépendants et indépendants ne se retrouvent pas regroupés ensemble.

\subsection{Tri par code de Gray}




Dans cette série d'expérimentations, nous comparons les performances du tri par code de Gray 
avec celles du tri lexicographique sur les données synthétiques de distribution uniforme et de Zipf. 
Rappelons que le tri peut s'appliquer
aux faits ou lors de l'assignation des codes aux bitmaps de l'index.
Pour chaque jeu de données, nous avons généré des index bitmaps en utilisant un tri 
lexicographique (Lex) et un tri par code de Gray (Gray). Nous avons également utilisé un simple
 regroupement (Random-sort) sur la table de faits avec la commande Unix «~sort --random-sort~»~:
 cette commande regroupe tous les faits identiques en séquences continues, mais n'ordonne
 pas les faits entre eux. Il est possible d'implémenter Random-sort
 en temps linéaire par rapport au nombre de faits ($O(n)$) alors que le tri lexicographique
 est en  $\Omega(n \log n)$.  
Le tri des faits peut être combiné avec le tri des codes assignés aux bitmaps (voir la Section~\ref{sec:methodedetri}). 
Si les faits sont triés lexicographiquement, l'allocation peut se faire 
en utilisant le tri par code de Gray (Lex-Gray) ou le tri alphabétique (Lex-Alpha  ou tout simplement Lex). 
Si les faits sont en revanche triés par code de Gray, nous n'étudions que 
l'allocation alphabétique des codes aux bitmaps (Gray-Alpha ou tout simplement Gray).
Pour rendre compte de l'amélioration apportée par les différentes méthodes de tri, 
nous avons aussi généré des index bitmaps pour des faits ordonnés aléatoirement (Random-shuffle). 
Nous calculons ainsi les améliorations relatives en espace de stockage de chaque tri par rapport à Random-shuffle.  

Nous étudions en particulier les effets 
qu'ont sur l'indexation par bitmap 
le nombre de dimensions, le biais des données et la valeur $k$ de l'encodage $k$~of~$N$. 
Nous avons aussi utilisé Census-Income (voir le Tableau~\ref{tab:caractDataSet}) pour 
mesurer les mêmes performances sur des données réelles. 

\begin{figure*}[!t]
\centering
  \subfloat[Distribution uniforme]{\includegraphics[width=0.32\textwidth, angle=270]{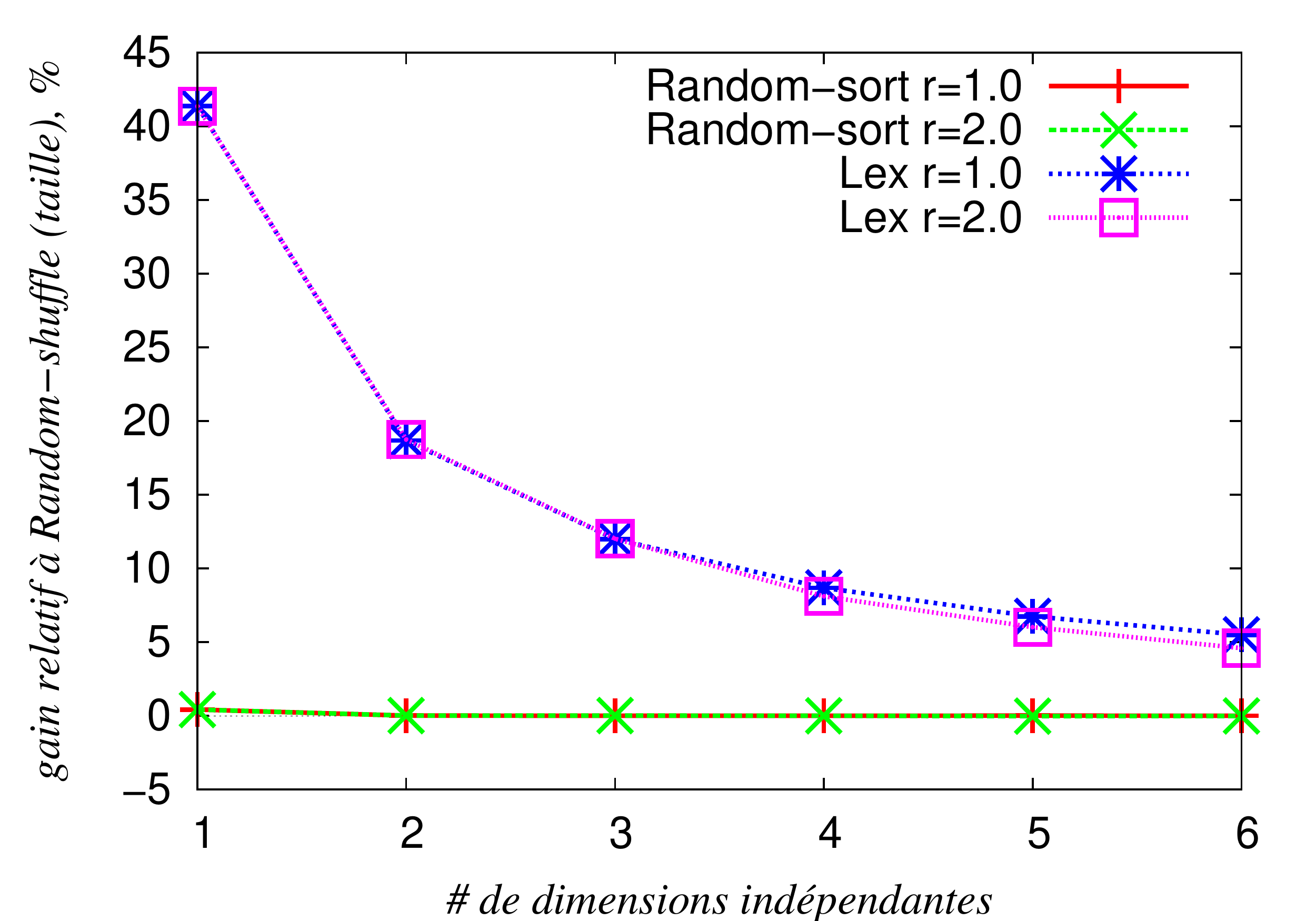}\label{fig:unidep-ktwo-rand}}
  \subfloat[Distribution de Zipf]{\includegraphics[width=0.32\textwidth, angle=270]{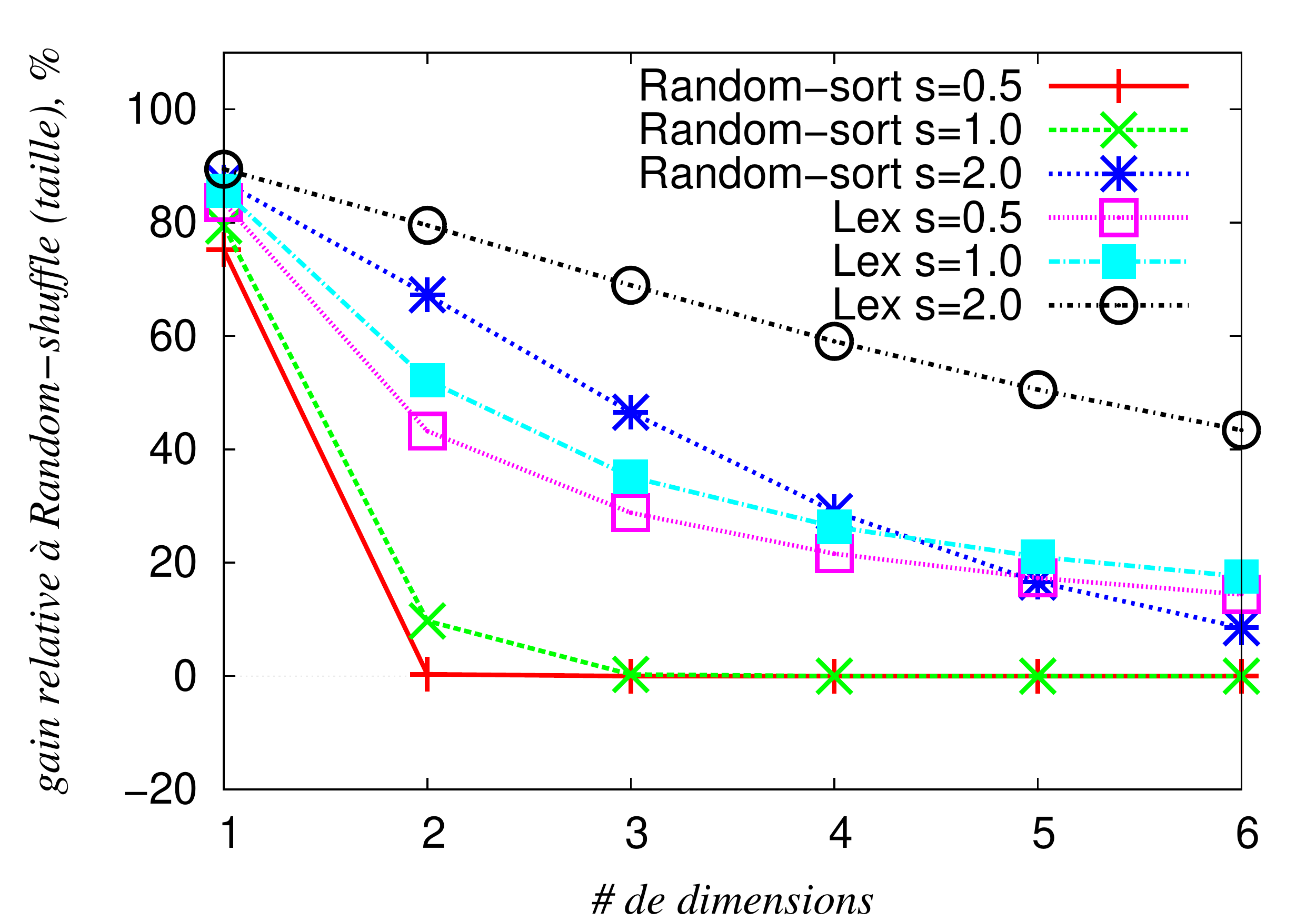} \label{fig:zipf-ktwo-rand}}
\caption{
Performance de Random-sort et Lex comparée à Random-shuffle en fonction de $d$, $k=2$. 
Pour les distributions uniformes, l'axe des $x$ donne le nombre de dimensions dépendantes. Les données ont en effet 2, 4, \ldots, 12 dimensions.
}
\end{figure*}

La Figure~\ref{fig:unidep-ktwo-rand} montre que 
le paramètre $r$, qui contrôle 
 la variété dans le nombre de valeurs distinctes, semble presque ne pas avoir du tout d'effet sur les différents jeux de données de distribution uniforme. 
 Le tri lexicographique réduit 
d'environ de moitié la taille de l'index lorsqu'il n'y a qu'une dimension
indépendante ($d=1$). 
Ce bénéfice est d'autant plus réduit que le nombre de dimensions augmente jusqu'à en devenir
négligeable. Par contre, Random-Sort semble totalement inefficace même lorsqu'il n'y a qu'une
dimension indépendante.

\begin{figure*} [!t]
\centering
 \subfloat[Distribution uniforme]{\includegraphics[width=0.32\textwidth, angle=270]{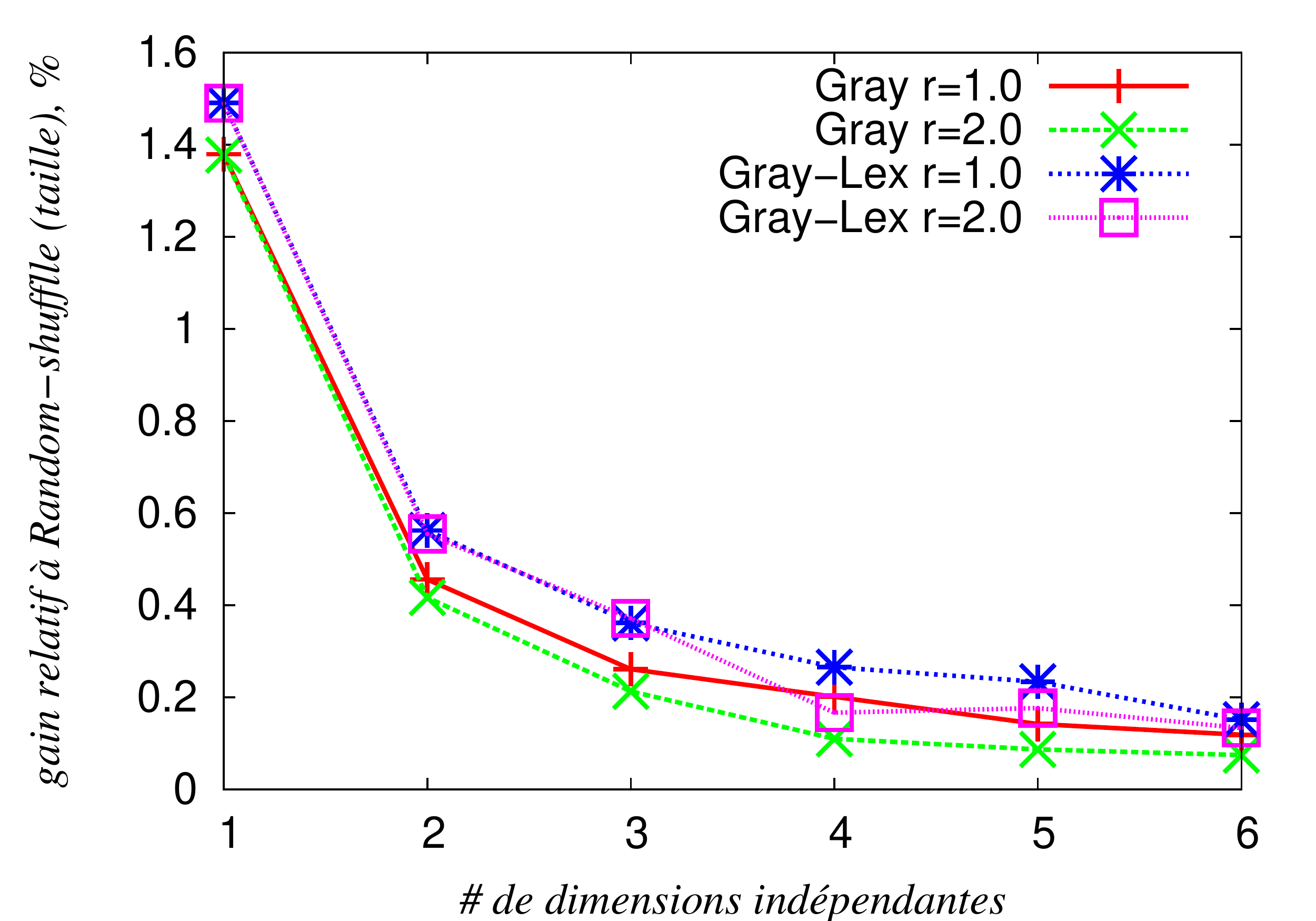}\label{fig:unidep-ktwo}}
 \subfloat[Distribution de Zipf]{\includegraphics[width=0.32\textwidth, angle=270]{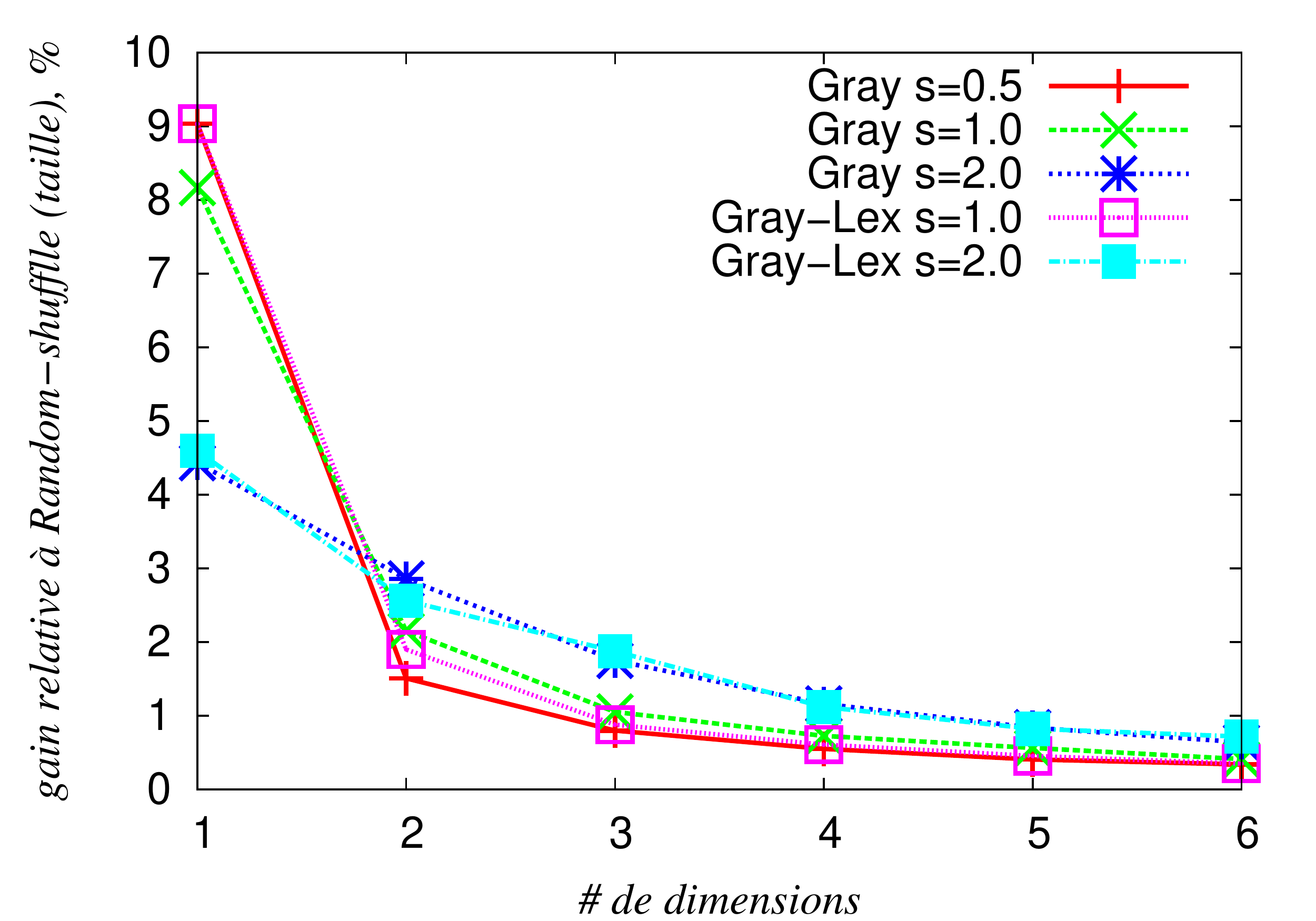}\label{fig:zipf-ktwo}}
\caption{
	Performance de Gray et Lex-Gray comparée au tri lexicographique en fonction de $d$, $k=2$.
}
\end{figure*}

La Figure~\ref{fig:unidep-ktwo} montre, pour différentes valeurs de $r$ de la distribution uniforme, 
l'impact que peut avoir le nombre de dimensions indépendantes $d$  sur les performances relatives
du  tri par code Gray et du tri lexicographique. Nous notons une amélioration 
marginale (moins de 1,5\,\%) 
de la taille de l'index lorsqu'il n'y a qu'une dimension indépendante. Cette amélioration diminue lorsque 
le nombre de dimensions et la valeur de $r$ augmentent. 
En comparant les deux versions du tri par code de Gray, 
nous constatons que le tri 
des codes alloués aux bitmaps 
 en tant que codes de Gray
 surpasse légèrement le tri par code Gray des faits.

À partir de la Figure~\ref{fig:zipf-ktwo-rand}, nous notons que le regroupement (Random-sort)
réduit la taille d'un index bitmap tout aussi bien que le tri lexicographique à condition qu'il n'y
ait pas plus d'une dimension au total.
Le tri lexicographique améliore davantage ces performances pour les distributions de Zipf  
d'au moins deux dimensions.
Cette figure montre bien que le tri lexicographique apporte un bénéfice substantiel, même pour 
un nombre de dimensions élevé, et que ce bénéfice croît en fonction du biais $s$.

La Figure~\ref{fig:zipf-ktwo} montre l'effet que peut avoir le nombre de dimensions sur 
les performances du tri par code de Gray par rapport au tri lexicographique 
pour $k=2$ et différentes valeurs du biais $s$ des distributions de Zipf.
Nous constatons que plus la distribution est biaisée, plus  
le bénéfice apporté par le tri par code de Gray est grand.
Dans nos données synthétiques, un biais plus grand correspond à un nombre de valeurs d'attributs plus 
petit et donc, à des index bitmaps plus denses. 
Nous pensons que le code de Gray est plus compétitif sur des index bitmaps denses. 
Cependant, cet avantage diminue rapidement lorsque le nombre des dimensions croît.
Pour une seule dimension, l'amélioration des performances est de 8\,\% ; au delà de 3 dimensions,
elle est en dessous de  2\,\%. La différence observée entre Gray-Lex et Gray est négligeable (moins de 2\,\% sur
la Figure~\ref{fig:unidep-ktwo} et
moins de 0,5\,\% sur la Figure~\ref{fig:zipf-ktwo}), sauf pour les données unidimensionnelles,
où Gray-Lex est 1\,\% moins bon que Gray.  

Des résultats similaires ont été observés pour $k=3$ ou $k=4$. 
Les valeurs de $k$ plus grandes améliorent légèrement (environ 5\,\% pour $d=2$ et $s=2.0$) 
les performances du tri par code de Gray par rapport au tri lexicographique. 
Ceci est également expliqué par le fait que le tri par code de Gray est plus compétitif sur des index bitmaps denses.

En comparant les Figures~\ref{fig:unidep-ktwo}~et~\ref{fig:zipf-ktwo}, nous notons
que Gray (et donc Gray-Lex, dont la performance moyenne est presque identique à celle de Gray) fournit 
une amélioration par rapport à Lex. Cependant, cette amélioration décroît rapidement selon le nombre croissant 
des dimensions.

Nous avons aussi utilisé trois jeux de données pour 
voir quels sont les bitmaps les plus affectés par le tri.
Pour minimiser le nombre de bitmaps (pour des raisons de lisibilité), nous n'avons utilisé que 
$k=4$ car peu de bitmaps sont produits pour cet encodage. 
 Nous avons généré
un fichier à distribution de Zipf comportant 8~449 faits et ayant un
biais unitaire ($s=1$). Les colonnes utilisent 16, 15, 16, et 15 bitmaps
et ont respectivement 1~382, 1~355, 1~378 et 1~345 valeurs distinctes.
 Nous avons aussi généré un fichier à distribution
uniforme comportant 20~000~faits et 3 dimensions indépendantes avec
des cardinalités respectives de 100, 200 et 400, et 3 dimensions
dépendantes générées avec l'algorithme mentionné à Section~\ref{sec:dim}. 
Les colonnes comptent, dans l'ordre, 200, 100, 400, 526, 537 et 536 valeurs
distinctes et elles utilisent 10, 9, 12, 13, 13 et 13 bitmaps.
Finalement, nous avons retenu deux extraits de Census-Income~: Census-Income~A décrit à la Section 
~\ref{tab:carUSIncome} et Census-Income~B obtenu en remplaçant la dimension la plus dominante 
d3 de ce jeu (cardinalité de 99~800) par la dimension 
«~dividends from stocks~»
de cardinalité plus faible (égale à 1~478).
Les colonnes de Census-Incomne~A utilisent 9, 15 et 41 bitmaps. Celles de Census-Income~B utilisent
9, 15 et 16 bitmaps.

\begin{figure*}[!t]
\centering
\subfloat[Distribution de Zipf]{\includegraphics[width=.3\textwidth,angle=270]{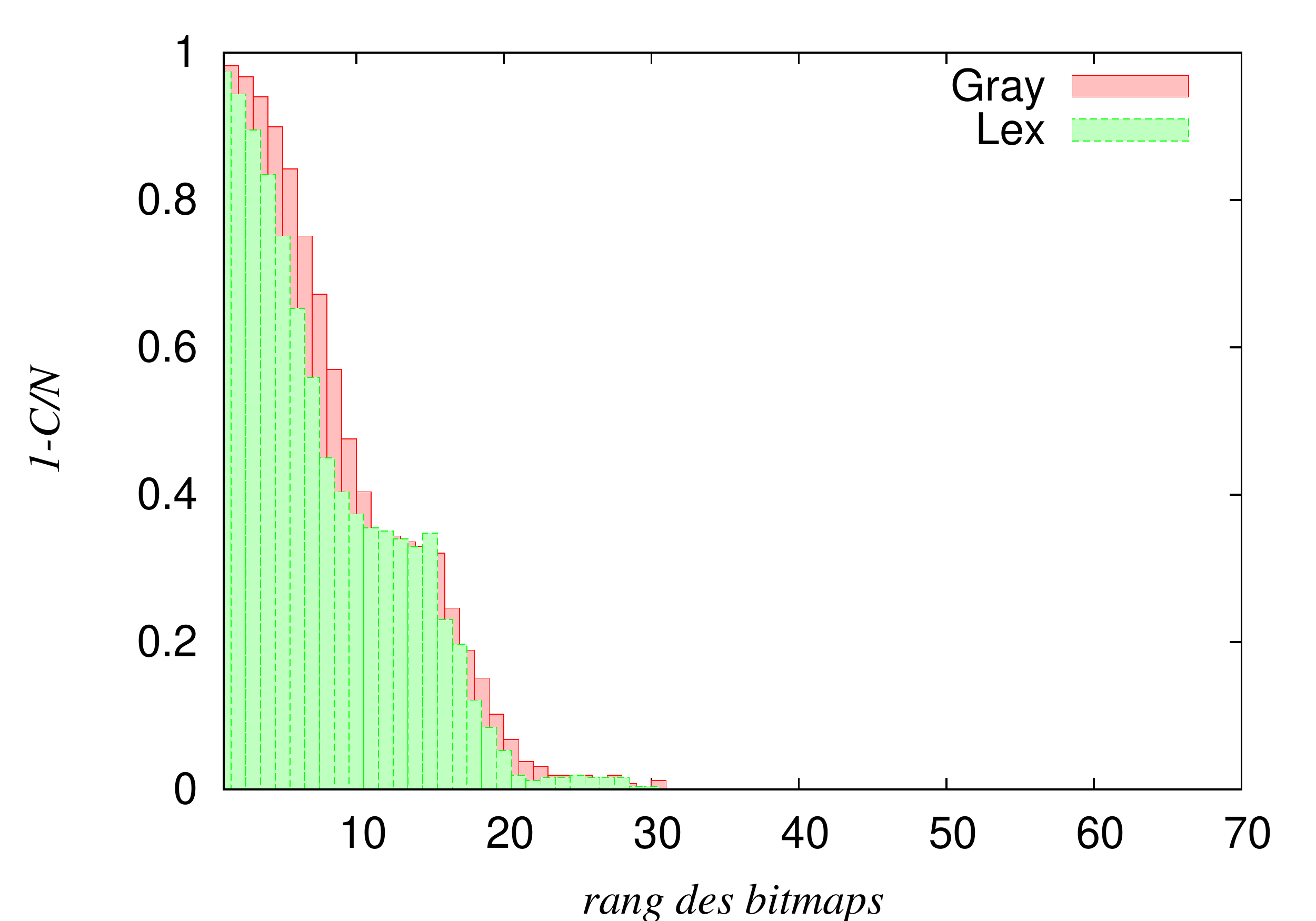}\label{fig:zipfhisto}}
\subfloat[Distribution uniforme]{\includegraphics[width=.3\textwidth,angle=270]{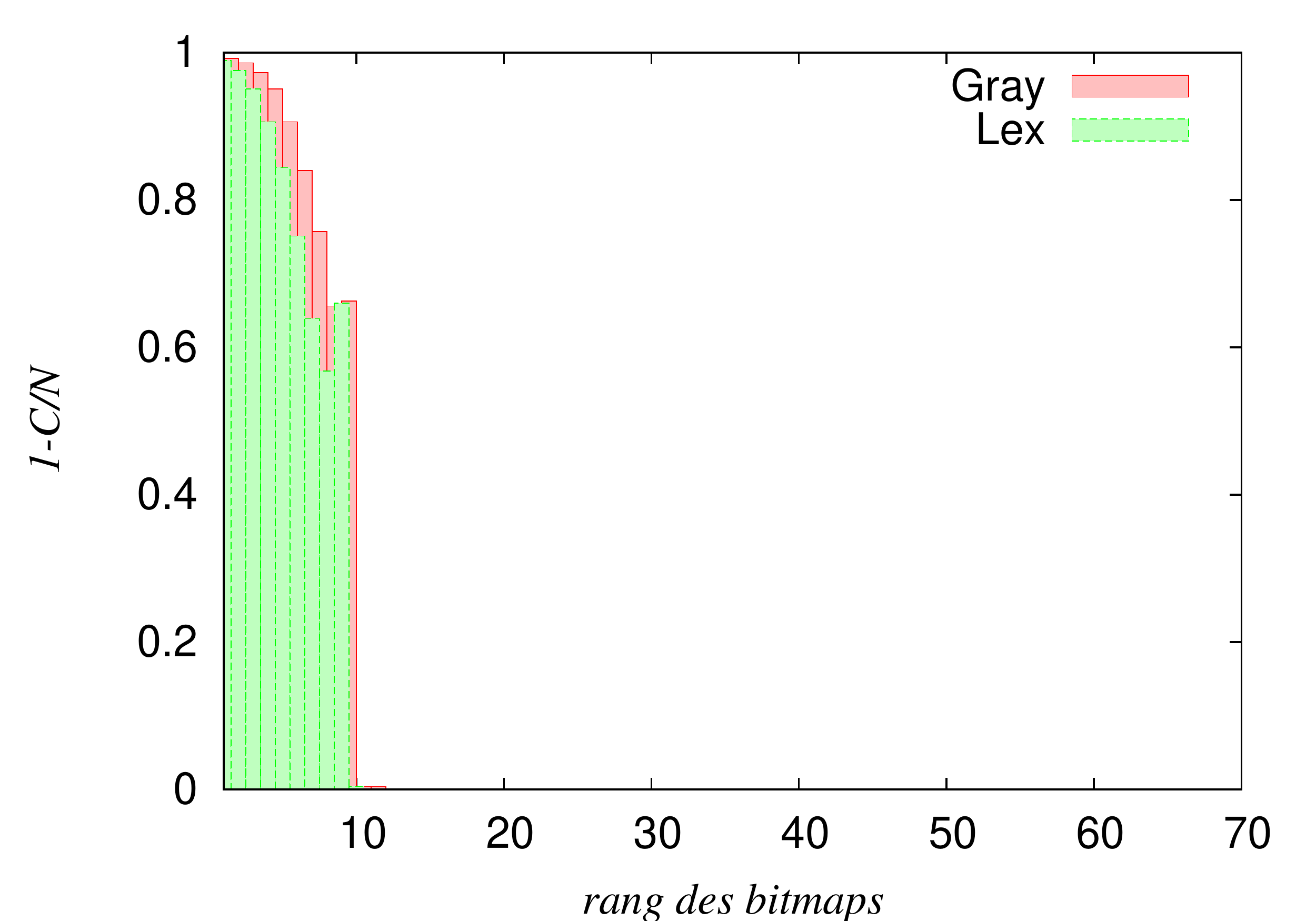}\label{fig:unidephisto}} \\
\subfloat[Census-Income~A]{\includegraphics[width=.3\textwidth,angle=270]{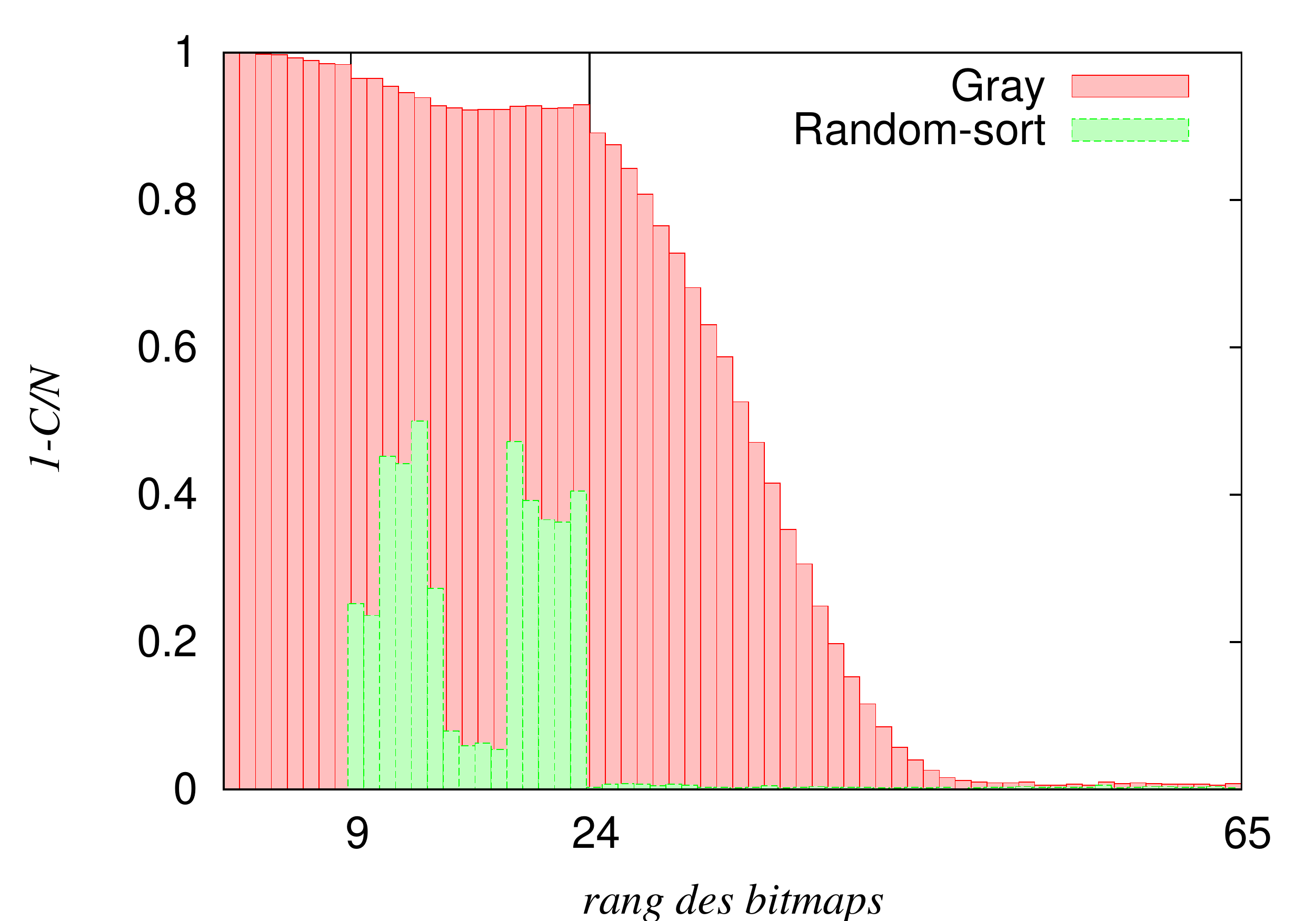}\label{fig:census-incomeAhisto}}
\subfloat[Census-Income~B]{\includegraphics[width=.3\textwidth,angle=270]{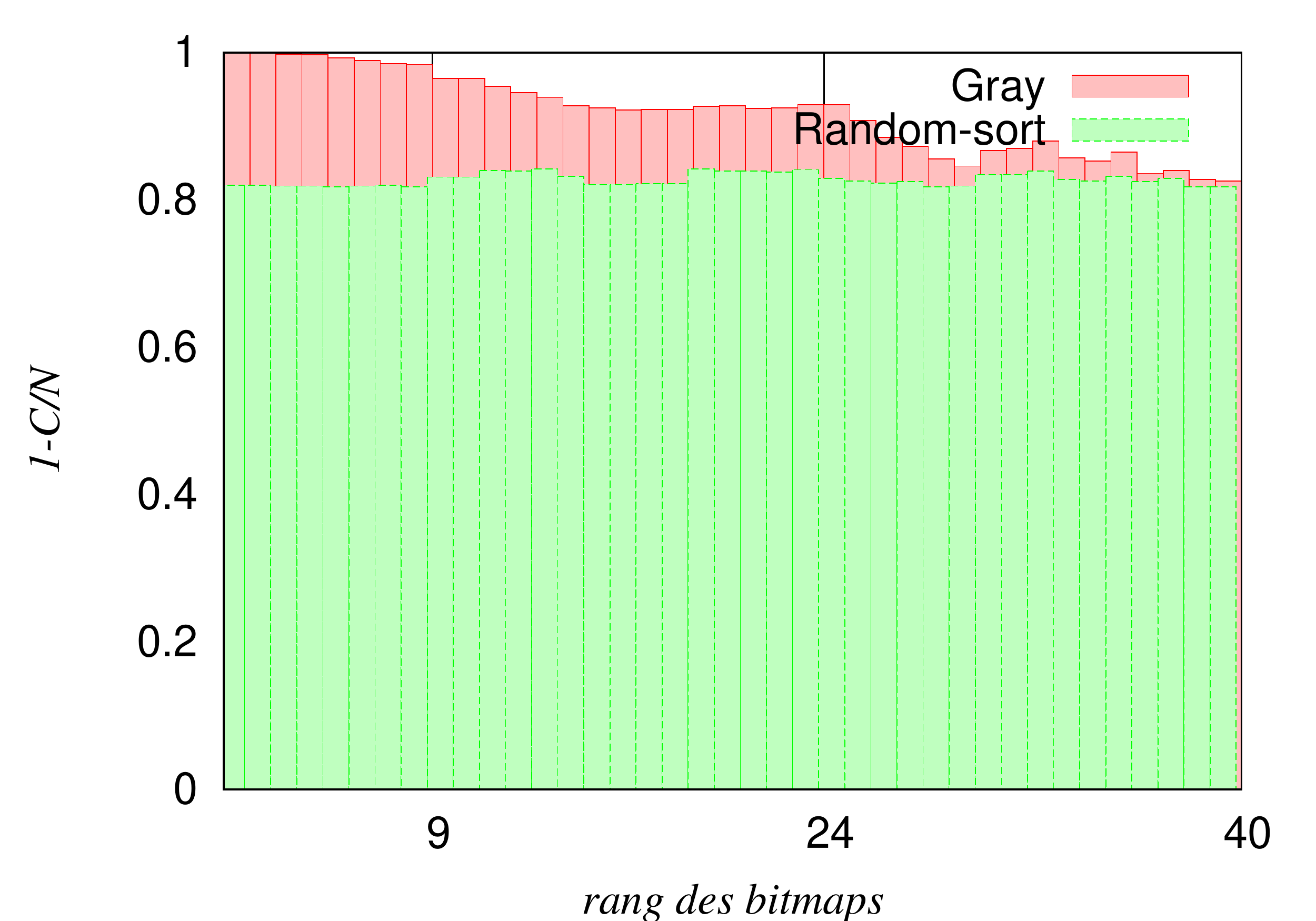}\label{fig:census-incomeBhisto}}
\caption{
	Tri par code de Gray et tri lexicographique des codes alloués aux bitmaps, $k=4$.
	$C$ est la taille après compression et $N$ la taille sans compression.
	Les bitmaps de tête sont mieux compressés.
	}
\end{figure*}

Dans les  Figures~\ref{fig:zipfhisto}~et~\ref{fig:unidephisto}, nous comparons l'effet de Gray et de Lex sur
les différents bitmaps. Le facteur mesuré ($1-C/N$) s'approche de 1 lorsque les bitmaps
sont fortement compressés, mais vaut 0 lorsqu'il y a peu de compression.
Les bitmaps les plus \og significatifs\fg{} de chaque dimension sont les plus à gauche.
 On
constate un phénomène qui peut être surprenant~: même si les bitmaps appartiennent
à différentes dimensions, le taux de compression des bitmaps décroît de manière monotone
 pour les données synthétiques (Figures~\ref{fig:zipfhisto} et \ref{fig:unidephisto}).
 On constate ce même phénomène pour Gray sur  Census-Income (Figure~\ref{fig:census-incomeBhisto}).
 Ce résultat suggère que le tri accélère les requêtes
 sur les premières colonnes davantage que sur les dernières, mais il accélère aussi les
 requêtes
 sur certaines valeurs plutôt que d'autres au sein d'une même colonne,
 selon que les bitmaps correspondants soient plus ou moins significatifs pour le tri.
 Par exemple, si l'allocation des bitmaps se fait de manière alphabétique, une requête
 portant sur la valeur Aaron pourra être plus rapide qu'une requête portant sur la valeur Zebra.
 
 
 Dans les Figures~\ref{fig:census-incomeAhisto}~et~\ref{fig:census-incomeBhisto}, nous avons comparé Random-Sort et Gray
 (Lex étant comparable à Gray).
 Alors que Gray compresse mieux que Random-Sort, on remarque aussi que Random-Sort 
 ne favorise pas la compression des bitmaps les plus à gauche. Ce résultat est prévisible
 si on considère que Random-Sort ne trie pas les valeurs.
 Alors que 
Census-Income~A et Census-Income~B ne diffèrent que du choix de la dernière colonne, 
Random-Sort a presque la même performance que Gray sur 
Census-Income~B alors que sa performance est bien moindre sur Census-Income~A.
Cela s'explique par la colonne d3 utilisée dans Census-Income~B qui a la valeur~0
avec une probabilité de 89\%. En effet, comme Random-Sort ne fait que regrouper
les faits identiques, il se trouve favorisé par une colonne ayant des valeurs relativement très fréquentes.
 
\cut{ 
À la première dimension l'allocation des codes par code de Gray surpasse l'allocation par ordre alphabétique, 
pour les bitmaps se trouvant au centre. L'allocation par Gray et alphabétique ont des résultats similaires pour la 
deuxième dimension, et les deux n'ont qu'un seul bitmap avec une compression notable sur la troisième dimension. 
L'allocation alphabétique a un avantage significatif sur la dernière dimension, mais ne dépasse pas le bénéfice 
apportée par l'allocation par code de Gray sur la troisième dimension.  } 
%
%
 
 \cut{
Le tri aléatoire (pas représenté) a un seul bitmap qui apporte une amélioration pour la première
dimension et un autre pour la troisième dimension. À partir de la Figure~\ref{fig:unidephisto}, nous pouvons voir
un effet similaire concernant la première dimension en utilisant l'allocation par Gray 
sur des données de distribution uniforme et dont les attributs sont dépendants. Une légère compression 
est observée sur les dimensions qui suivent la première. De plus, aucun bitmap alloué par un tri 
aléatoire n'apporte d'amélioration. }
 

\cut{
À la Figure~\ref{fig:tweedhisto}, nous pouvons voir que le meilleur ordonnancement bénéficie à la 
première dimension (le premier groupe de bitmaps), mais continue d'avoir de bonnes performances
pour de nombreux bitmaps appartenant à le deuxième et à la troisième dimension. Le tri aléatoire apporte
40\,\% d'entrées propres pour presque tous les bitmaps. 
} 

\cut{
Random-sort had about 40\% clean entries for most of its bitmaps.
Why?
\kamel{Owen, do you have any answer?}
\owen{I don't have a histogram to give precise numbers.  However
this TWEED projection has on average 5 duplicate of each unique record
(there are about 1900 unique records and 11000 records).  Looking 
at the data, though, we see that there are many duplicates of a 
few popular records.  I think the skewed popularity of these few 
records explains most of this.  Plus, if you have only one transition
from one attribute value to another, with a reasonably sparse code,
for any given bitmap the transition is fairly likely to be from
a 0 (in the first code) to another 0 (in the second code).  This
dataset was coded for $k=4$, but actually uses 3-of-8 encoding
for 53 values, 2-of-7 encoding for 16 values, 2-of-6 encoding for 11
values, and 4-of-11 coding for 287 values.  About 2/3 of the bits in the
codes are zeros.}
\kamel{Daniel, que penses-tu de supprimer tout ce qui est en relation avec TWEED et random sort ? Ça alourdit le papier et rend sa lecture un peu confuse.}
\daniel{Je pense que l'intention d'Owen est effectivement qu'on remplace TWEED par autre chose.}
\owen{TWEED should be replaced by some projection of one of the
real data sets we use, so we don't unnecessarily introduce another
data set.  But don't use DBGEN.  It is fake and I don't trust it.
I recommend CensusIncome since it has lots of dimensions.
If the columns of the chosen data set are ordered by increasing or decreasing
cardinality (as suggested by DL's comment in the next section)
and if we use -a gray, then a figure like fig\ref{fig:tweedhisto}
can help show the effect of adding more and more dimensions.
Rather than show the individual bitmaps, we could have one column
for each dimension, showing the compression of that dimension.
Maybe the dimensions would be shown on the x axis in order of
cardinality.  We could
stop comparing alpha vs Gray, but for simplicity, just show the Gray
result.  It would not really fit in with this section's goal of comparing
alpha and gray, but it leads into the next section.\\
But I would also like (if there is room)
to see the individual bitmaps for the
first 4-5 dimensions, much like the current TWEED plot.  It is
so different from the synthetic data.  }
}

\subsection{Effet de  l'ordre des colonnes}



Le tri de la table de faits nécessite un ordonnancement des colonnes.
Puisque le tri se fait avant la construction de l'index, seuls les histogrammes
des valeurs par dimension sont connus. Deux stratégies simples s'offrent
à l'analyste : trier en premier sur les petites dimensions ou sur les dimensions
dont la cardinalité est plus importante. Les deux techniques ont leurs avantages.
 Les dimensions de cardinalité élevée générèrent plus de bitmaps et peuvent occuper 
plus d'espace si elles sont mal ordonnées. Dans la mesure où plusieurs de leurs valeurs sont relativement fréquentes (plus de 32~occurrences), un tri privilégiant une grande dimension est donc plus avantageux.
 En plaçant une dimension ayant un grand nombre de valeurs distinctes en premier, 
les dimensions subséquentes seront relativement peu triées : on trouvera peu de plages de valeurs constantes.
 Nous pouvons tirer la conclusion suivante~: si la cardinalité d'une des dimensions est beaucoup 
plus importante que celle des autres et comprend plusieurs valeurs dont l'occurrence est élevée, il est alors plus judicieux de la mettre en tête du tri. En revanche, s'il y a peu de différences entre les cardinalités, il est peut-être plus préférable de trier d'abord à partir des dimensions de faible cardinalité afin que l'ensemble des colonnes puissent bénéficier du tri.

\begin{figure*}[!t]
\centering
  \subfloat[Census-Income]{\includegraphics[width=0.4\textwidth]{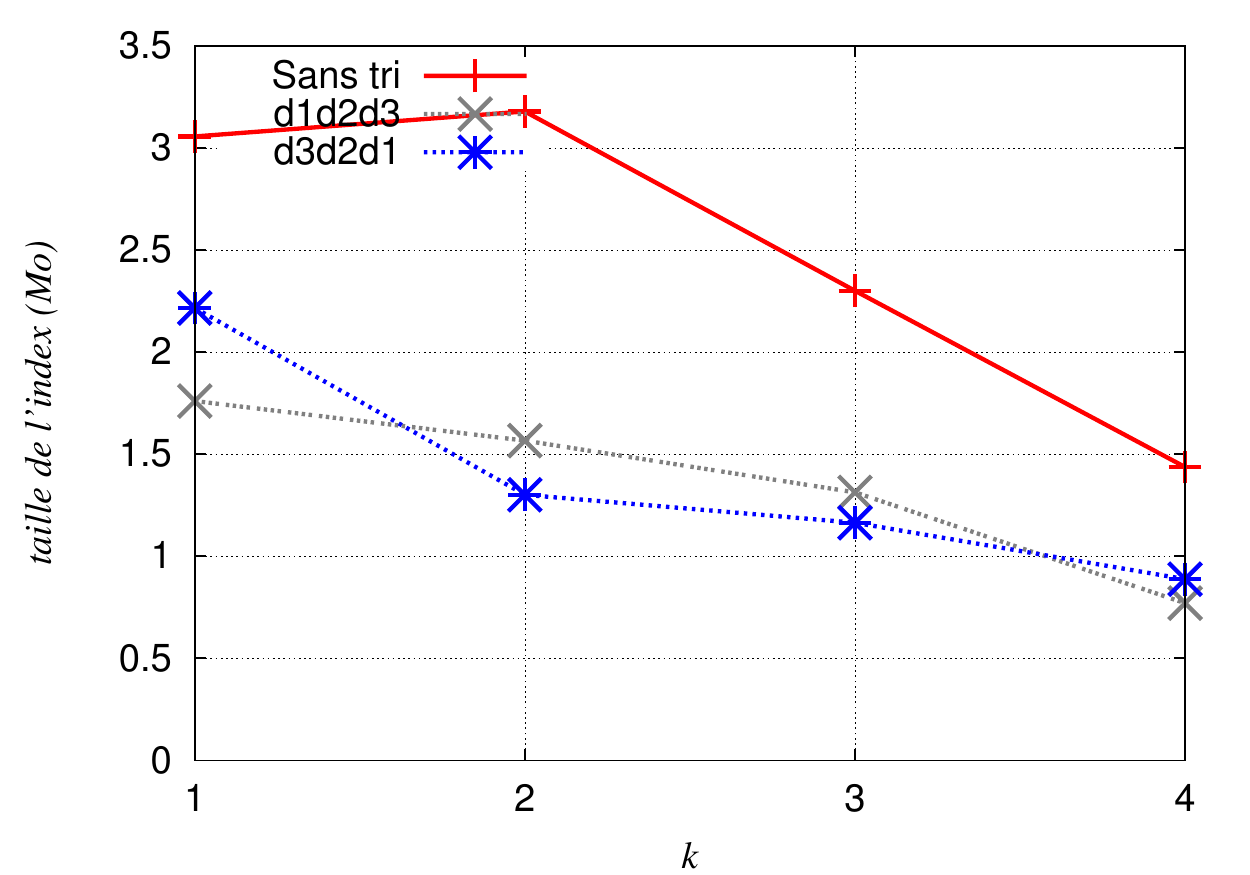}}
  \subfloat[DBGEN]{\includegraphics[width=0.4\textwidth]{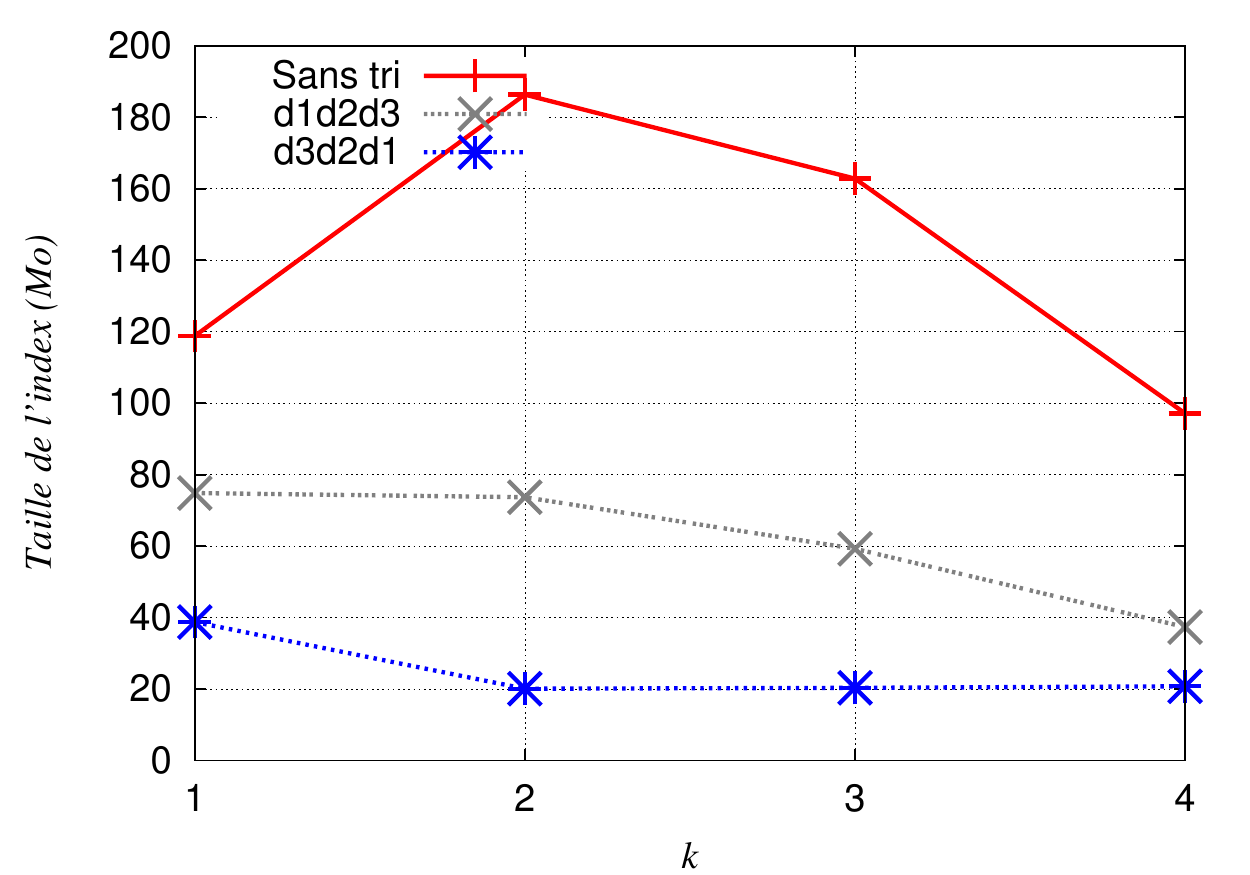}} \\
\subfloat[DBLP]{\includegraphics[width=0.4\textwidth]{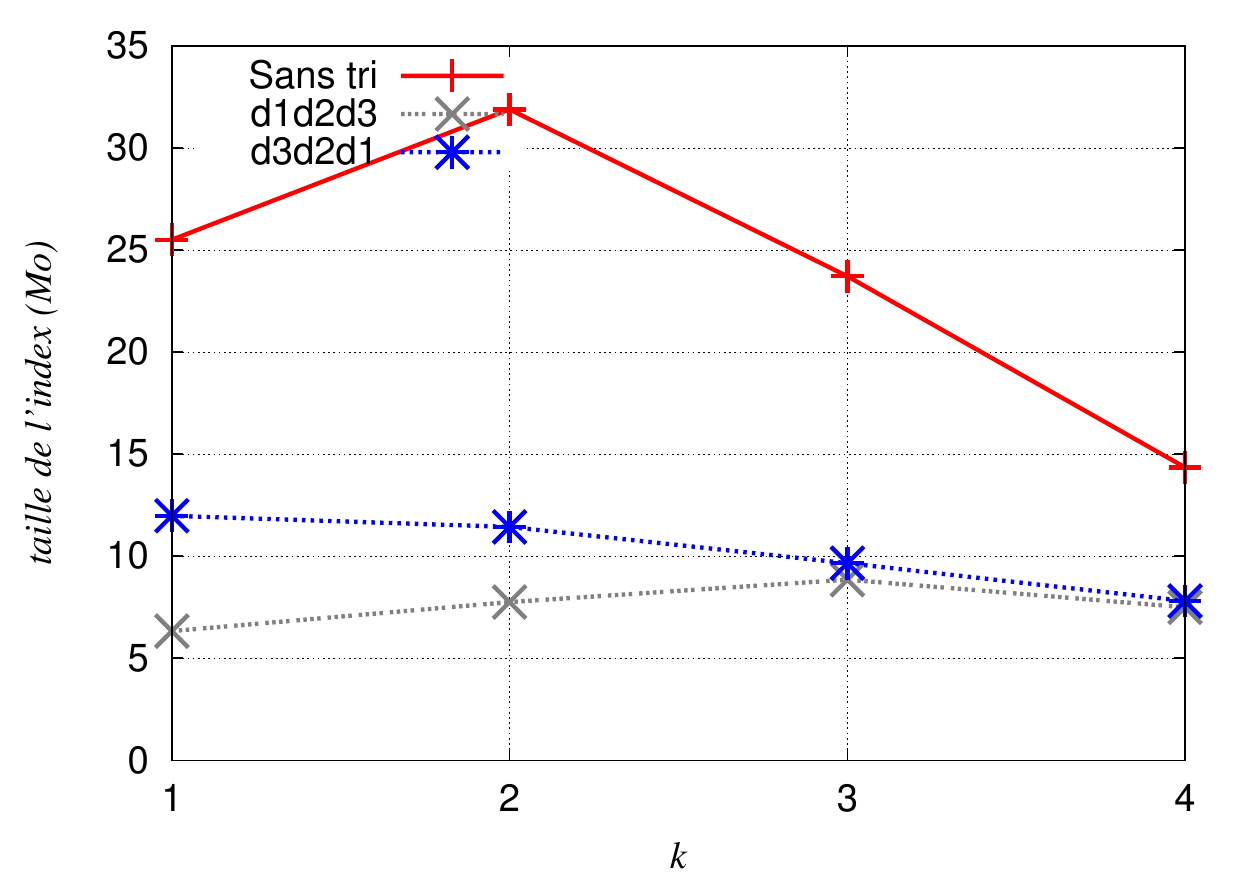}} 
  \subfloat[Netflix]{\includegraphics[width=0.4\textwidth]{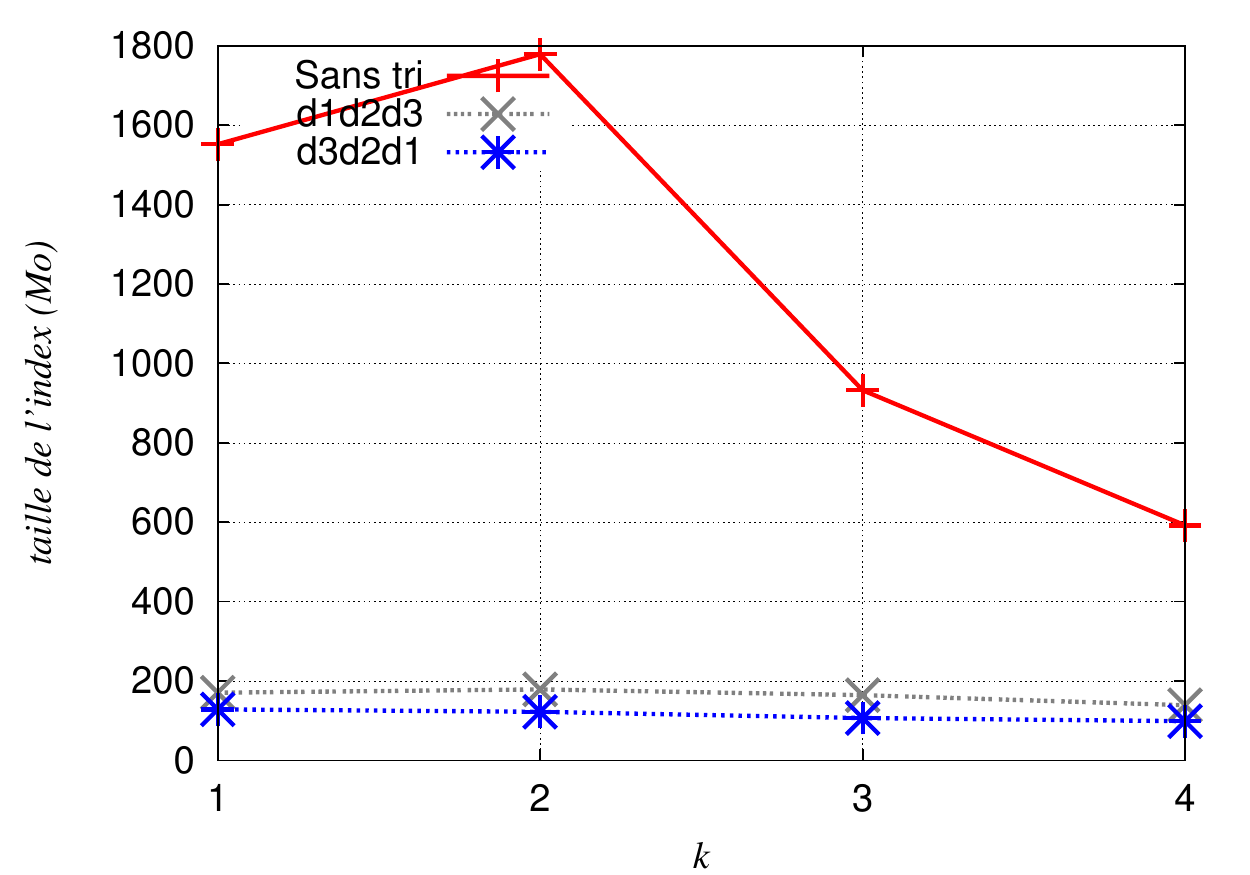}}
\caption{Taille de l'index selon que les données ne soit pas triées ou triées lexicographiquement à partir
de la dimension la plus importante (d3d2d1) ou de la dimension la plus petite.\label{fig:sizesortcolumnorder}}
\end{figure*}

La Figure~\ref{fig:sizesortcolumnorder} montre une comparaison de l'efficacité de deux
ordonnancement des colonnes. Pour ces tests, nous n'avons retenu que 3~colonnes
pour chacune des tables. Les gains en compression dus au tri lexicographique sont d'au moins
20\%, mais peuvent atteindre un facteur de dix dans le cas de Netflix.
Nous constatons que l'effet de l'ordre des colonnes semble être moins
prononcé pour l'encodage 4-of-$N$ que pour l'encodage 1-of-$N$. Cela s'explique par le fait
que le nombre de bitmaps générés par une grande dimension est moindre pour l'encodage 4-of-$N$ que pour 1-of-$N$.
L'effet de l'ordonnancement des dimensions est relativement important. Pour DBGEN, il est
nettement préférable de trier selon la plus grande dimension alors que pour DBLP, il est nettement
préférable de trier selon la plus petite dimension. 
II suffit de consulter le Tableau~\ref{tab:carUSIncome} pour se rendre compte  qu'il n'est pas préférable 
de trier selon la plus grande dimension pour DBLP~: elle comporte un demi-million de valeurs distinctes alors que la table de faits n'en comporte que 1,5 million de faits. Ce qui représente une fréquence moyenne inférieure à 3 pour chaque valeur distincte. Pour Netflix et Census-Income, la différence
entre les deux ordonnancements est plus modérée (environ 30\,\%). Le Tableau~\ref{tab:sizesortcolumnorder}
donne le détail de l'effet du tri sur les différentes dimensions selon l'ordre des colonnes.

\begin{table*}[!t]
\begin{center}
\begin{scriptsize}
\begin{tabular}{|c|rrr|rrr|}\cline{2-7}
	\multicolumn{1}{c|}{}  & \multicolumn{3}{c|}{\textbf{Census-Income}}  & \multicolumn{3}{c|}{\textbf{DBGEN}} \\ \cline{2-7}
	\multicolumn{1}{c|}{} & sans tri &\multicolumn{1}{c}{d1d2d3} & \multicolumn{1}{c|}{d3d2d1} & sans tri & \multicolumn{1}{c}{d1d2d3} & \multicolumn{1}{c|}{d3d2d1} \\ \hline
	d1 & 0,27$\times 10^6$  & 448 & 0,25$\times 10^6$  &   2,58$\times 10^6$ & 75     &   2,58$\times 10^6$   \\
	d2 & 29~176 & 12~036 &  28~673 & 4,12$\times 10^6$ & 347     &   4,11$\times 10^6$ \\
	d3 & 0,50$\times 10^6$ & 0,45$\times 10^6$ & 0,30$\times 10^6$ &  24,45$\times 10^6$  & 19,62$\times 10^6$      &   3,46$\times 10^6$  \\ \hline
	total & 0,80$\times 10^6$ & 0,46$\times 10^6$ & 0,58$\times 10^6$ &  31,15$\times 10^6$ & 19,62$\times 10^6$     &  10,16$\times 10^6$  \\ \hline
\end{tabular}

\begin{tabular}{|c|rrr|rrr|}\cline{2-7}
	\multicolumn{1}{c|}{}  & \multicolumn{3}{c|}{\textbf{DBLP}} &\multicolumn{3}{c|}{\textbf{Netflix}} \\ \cline{2-7}
	\multicolumn{1}{c|}{} & sans tri &\multicolumn{1}{c}{d1d2d3} & \multicolumn{1}{c|}{d3d2d1} & sans tri & \multicolumn{1}{c}{d1d2d3} & \multicolumn{1}{c|}{d3d2d1} \\ \hline
	d1 & 0,79$\times 10^6$ & 237 & 0,55$\times 10^6$  &  15,54$\times 10^6$ & 256 &   12,47$\times 10^6$        \\
	d2 & 2,62$\times 10^6$ & 42~407 &  0,97$\times 10^6$  & 0,19$\times 10^9$ &   0,14$\times 10^6$       &   20,30$\times 10^6$        \\
	d3 & 3,27$\times 10^6$ & 1,61$\times 10^6$ &  1,61$\times 10^6$      &   0,20$\times 10^9$ &  44,52$\times 10^6$    &    0,92$\times 10^6$       \\ \hline
	total & 6,68$\times 10^6$ &  1,61$\times 10^6$ &  2,13$\times 10^6$      &   0,39$\times 10^9$ &  44,66$\times 10^6$      &      33,69$\times 10^6$     \\ \hline
\end{tabular}
\end{scriptsize}
\end{center}
\caption{Nombre de mots de 32 bits utilisés pour les différents bitmaps selon que les données
aient été triées lexicographiquement à partir
de la dimension la plus importante (d3d2d1) ou la dimension la plus petite (d1d2d3) pour $k=1$.
}\label{tab:sizesortcolumnorder}
\end{table*}

Le tri tend à créer des plages de valeurs identiques sur les premières colonnes.
Les effets bénéfiques  du tri sont sans doute moins apparents sur les dernières 
colonnes (voir le Tableau~\ref{tab:sizesortcolumnorder}),
à moins que celles-ci soient fortement corrélées avec les premières. Si le nombre
de colonnes est élevé, les bénéfices du tri peuvent être moindres et voire même s'estomper.
Dans le Tableau~\ref{tab:sizesortcolumnorderd10}, nous avons trié des projections de Census-Income et
DBGEN sur 10~dimensions $d1 \ldots d10$ ($|d1| < \ldots  < |d10|$, où $|x|$ désigne la cardinalité de la dimension $x$). 
Les dimensions $d1 \dots d3$ faisant parties de ces 10~dimensions ne correspondent 
pas à celles illustrées au Tableau~\ref{tab:carUSIncome}.  
Nous notons que pour Census-Income, peu importe
l'ordre des colonnes, l'effet du tri se fait sentir jusqu'à la dernière colonne. 
Pour DBGEN, l'effet se fait sentir sur toutes les colonnes si on tri d'abord sur la
colonne ayant la plus petite cardinalité, alors que l'effet ne persiste que sur 5~colonnes
lorsque l'on tri à partir de la plus grande cardinalité (d10). Dans tous les cas,
l'effet du tri est nettement plus présent sur les premières dimensions triées que
sur les dernières.
 
Si on compare le Tableau~\ref{tab:sizesortcolumnorder} avec le Tableau~\ref{tab:sizesortcolumnorderd10},
on constate que pour Census-Income, il importe peu qu'on trie à partir de la
colonne de plus faible cardinalité (d1) ou de plus forte cardinalité.
Cela s'explique 
parce que nous avons retenu une colonne de très forte
cardinalité (99~800~valeurs distinctes) par rapport au nombre total de faits (199~523).
Pour DBGEN, l'avantage du tri à partir de la colonne de plus forte cardinalité est moindre
avec dix~colonnes (Tableau~\ref{tab:sizesortcolumnorderd10}) qu'avec trois~colonnes (Tableau~\ref{tab:sizesortcolumnorder}).
Cela s'explique par le fait que dans le premier cas, nous avons introduit la colonne d10 qui 
comporte un million de faits distincts contre un total
de près de 14~millions de faits pour DBGEN alors que la colonne d3 dans le Tableau~\ref{tab:sizesortcolumnorder}
ne comporte que 400~milles valeurs distinctes.

\cut{
\daniel{Kamel, ici, on a un problème. Compare le Tableau~\ref{tab:sizesortcolumnorder}avec le Tableau~\ref{tab:sizesortcolumnorderd10}, les conclusions sont pratiquement inversées. Pour Census-Income
d1d2d3 est nettement préférable, alors que maintenant $d10 \ldots d1$ est nettement mieux. Pour DBGEN,
il n'y a plus de différences entre les deux alors qu'on $d3d2d1$ était nettement meilleur auparavant.
Faudrait qu'on puisse expliquer ce phénomène en vertu de nos lignes directrices énoncées plus haut.
Est-ce que tu peux vérifier? }
\kamel{J'ai fait des vérifications et je confirme la tendance. Ceci s'explique par le fati que le dimension choisie d3 de Census-Income 
a une cardinalité très forte par rapport au nombre total de faits (99800/199523). L'effet du tri sur cette dimension est  atténué du fait de sa grande cardinalité relative. J'ai effectué un autre test, toujours avec 3  dimensions mais en replaçant cette dimension et le tri $d3d2d1$  est meilleur : les dimensions utilisées sont 0, 5 et 18 et les totaux de la taille sont 349884 (sans tri), 43789 (d1d2d3) et 33026 (d3d2d1).}
\daniel{La confusion dans mon esprit vient du fait que je présumais que $d1,d2,\ldots, d10$ incluaient d1, d2, et d3.
Si c'était vrai, alors la cardinalité de d10 serait plus grande que d3, ce qui n'est pas le cas si je comprend bien?}
\daniel{On pourrait facilement régler cette confusion si avait les cardinalités des différentes dimensions. Enfin, au moins
pour les cardinalités maximales.}
}
\begin{table*}[!t]
\centering
\begin{scriptsize}
\begin{tabular}{|c|rrrr|rrrr|}\cline{2-9}
	\multicolumn{1}{c|}{}  & \multicolumn{4}{c|}{\textbf{Census-Income}}  & \multicolumn{4}{c|}{\textbf{DBGEN}} \\ 
	\multicolumn{1}{c|}{} & cardinalités &sans tri &\multicolumn{1}{c}{d1 \ldots d10} & \multicolumn{1}{c|}{d10 \ldots d1} & cardinalités & sans tri & \multicolumn{1}{c}{d1 \ldots d10} & \multicolumn{1}{c|}{d10 \ldots d1} \\ \hline
d1 & 7 & 42~427 & 32 & 42~309 & 2 & 0,75$\times 10^6$ & 24 & 0,75$\times 10^6$ \\
d2 & 8 & 36~980 & 200 & 36~521 & 3 &  1,11$\times 10^6$ & 38 & 1,11$\times 10^6$ \\
d3 & 10 & 34~257 & 1~215 & 28~975 & 7 & 2,58$\times 10^6$ & 150 & 2,78$\times 10^6$ \\
d4 & 47 & 0,13$\times 10^6$ & 12~118 & 0,13$\times 10^6$ & 9 & 0,37$\times 10^6$ & 100~6 & 3,37$\times 10^6$ \\
d5 & 51 & 35~203 & 17~789 & 28~803 & 11 & 4,11$\times 10^6$ & 10~824 & 4,11$\times 10^6$ \\
d6 & 91 & 0,27$\times 10^6$ & 75~065 & 0,25$\times 10^6$ & 50 & 13,60$\times 10^6$ & 0,44$\times 10^6$ & 1,42$\times 10^6$ \\
d7 & 113 & 12~199 &  9~217 & 12~178 & 2~526 & 23,69$\times 10^6$ & 22,41$\times 10^6$ & 23,69$\times 10^6$ \\
d8 & 132 & 20~028 & 14~062 & 19~917 & 20~000 & 24.00$\times 10^6$ & 24,00$\times 10^6$ & 22,12$\times 10^6$ \\
d9 & 1~240 & 29~223 & 24~313 & 28~673 & 400~000 & 24,84$\times 10^6$ & 24,84$\times 10^6$ & 19,14$\times 10^6$ \\
d10 & 99~800 & 0,50$\times 10^6$ & 0,48$\times 10^6$ & 0,30$\times 10^6$ & 984~298 & 27,36$\times 10^6$ & 27,31$\times 10^6$ & 0,88$\times 10^6$ \\ \hline
total  & - & 1,11$\times 10^6$ & 0,64$\times 10^6$ & 0,87$\times 10^6$  & - & 0,122$\times 10^9$ & 0,099$\times 10^9$ & $0,079\times 10^9$ \\ \hline
\end{tabular}
\end{scriptsize}
\caption{Nombre de mots de 32 bits utilisés pour les différents bitmaps selon que les données
soient triées lexicographiquement à partir
de la dimension la plus importante (d10\ldots d1) ou la dimension la plus petite (d1\ldots d10) pour $k=1$. 
}\label{tab:sizesortcolumnorderd10}
\end{table*}

\kamelcut{
\begin{table*}
\centering
\begin{scriptsize}
\begin{tabular}{|c|rrr|rrr|rrr|rrr|}\cline{2-13}
	\multicolumn{1}{c|}{}  & \multicolumn{3}{c|}{Census-Income}  & \multicolumn{3}{c|}{DBGEN} & \multicolumn{3}{c|}{DBLP} &\multicolumn{3}{c|}{Netflix} \\ 
	\multicolumn{1}{c|}{} & sans tri &\multicolumn{1}{c}{d1d2d3} & \multicolumn{1}{c|}{d3d2d1} & sans tri & \multicolumn{1}{c}{d1d2d3} & \multicolumn{1}{c|}{d3d2d1} & sans tri & \multicolumn{1}{c}{d1d2d3} & \multicolumn{1}{c|}{d3d2d1} & sans tri &  \multicolumn{1}{c}{d1d2d3} &    \multicolumn{1}{c|}{d3d2d1} \\ \hline
	d1 & 2,6$\times 10^6$ & 448 & 0,3$\times 10^6$  &   - & 72     &   3,5$\times 10^6$  & - & 237 & 0,5$\times 10^6$  &  - & 256 &   12,5$\times 10^6$        \\
	d2 & 4,1$\times 10^6$ & 12,036 &  28,673 &   - & 342     &   4,1$\times 10^6$  & - & 42407 &  1,0$\times 10^6$  & - &   0,1$\times 10^6$       &   20,3$\times 10^6$        \\
	d3 & 2,4$\times 10^6$ & 0,4$\times 10^6$ & 0,3$\times 10^6$ &  - & 19,6$\times 10^6$      &   2,6 $\times 10^6$  & - & 1,6$\times 10^6$ &  1,6$\times 10^6$      &   - &  44,5$\times 10^6$    &    0,9$\times 10^6$       \\ \hline
	total & 7,1$\times 10^6$ & 0,4$\times 10^6$ & 0,6$\times 10^6$ &  - & 19,6$\times 10^6$     &  10,0$\times 10^6$  & - &  1,6$\times 10^6$ &  3,1$\times 10^6$      &   - &  44,6$\times 10^6$      &      33,7$\times 10^6$     \\ \hline
\end{tabular}
\caption{Nombre de mots de 32 bits utilisés pour les différents bitmaps selon que les données
soient triées lexicographiquement à partir
de la dimension la plus importante (d3d2d1) ou la dimension la plus petite (d1d2d3) pour $k=1$.}\label{tab:sizesortcolumnorder}
\end{scriptsize}
\end{table*}

}

L'ordonnancement des colonnes a un effet sur les performances de l'index dans le sens où un index de plus petite taille suite à un tri sera plus rapide. Si le tri favorise une colonne plutôt qu'une
autre, les requêtes portant sur cette colonne devraient en bénéficier.

On pourrait aussi élaborer d'autres stratégies plus sophistiquées en tenant pleinement compte
des histogrammes. Par exemple, une dimension n'ayant que des valeurs avec une fréquence
inférieure à 32 ne devrait sans doute pas servir de base au tri. Dans le cas où nous
avons beaucoup de colonnes et qu'on souhaite tout de même tirer profit du
tri, on peut envisager des techniques comme le partionnement vertical.
Canahuate~\cite{pinarunpublished---fr} établissent non pas l'ordre des colonnes, mais aussi l'ordre
des bitmaps~: leur approche suppose que l'index bitmap a été préalablement matérialisé avant le tri. 

\subsection{Effet du tri par bloc}

Jusqu'à présent, nous n'avons considéré que le tri de la totalité de la table de faits.
Cette opération peut s'avérer coûteuse pour les grandes tables de faits.
En effet, le tri est une opération qui requiert un temps $\Omega(n \log n)$ alors
que l'indexation d'une table de faits ayant un nombre déterminé de valeurs par colonne
est une opération en temps $\Omega(n)$~: pour $n$ suffisamment grand, le
tri dominera le temps de construction des index. Un tri par bloc sans
fusion des blocs peut se faire en temps $O(n \log n/B)$ où $B$ est le nombre de blocs.
La différence entre $\log n$ et $\log n/B$ peut être significative. Si on considère
 $n/B$ comme une constante qui ne dépend que de la mémoire disponible, alors le temps
 du tri devient une opération en temps linéaire.

La question qui se pose est de savoir si le tri par bloc peut avoir des bénéfices comparables au tri complet.
 Pour tester cette hypothèse,
nous avons retenu nos deux plus grosses tables (DBGEN et Netflix)  et trié lexicographiquement. Nous avons projeté
chaque table sur trois dimensions de cardinalités différentes (voir la Section~\ref{sec:dim}). 
La taille d'un bloc est obtenue en divisant le nombre total de faits par le nombre de blocs souhaité. Cette valeur
est passée à la commande Unix «~split~» qui se charge de scinder la table de faits en plusieurs blocs, qui sont 
ensuite à leur tour triés par la commandes «~sort~». Ces blocs ainsi triés sont finalement fusionnés
en utilisant la commande Unix «~cat~» pour  obtenir
une table de faits triée par bloc. Le Tableau~\ref{tab:SortBlockTime} montre que le temps du tri par bloc diminue lorsque
le nombre de blocs augmente. Cela est dû au fait que lorsque le nombre de blocs est élevé, on a moins de données
à trier par bloc et donc un tri plus rapide.

\begin{table*}[!t]
\centering
\begin{scriptsize}
\begin{tabular}{|c|ccccc|ccccc|} \cline{2-11}
 \multicolumn{1}{c}{} &  \multicolumn{5}{|c|}{\textbf{DBGEN}} & \multicolumn{5}{|c|}{\textbf{Netflix}} \\ \hline 
\# de blocs	& tri & fusion  & indexation  &total  & taille & tri  & fusion & indexation & total & taille \\ \hline  
tri complet & 31 & - &  65 & 96 & 39  & 487 & - & 558 & 1~045  & 129\\
5 & 28 & 2 & 68 & 98  & 51  &	360	& 85 & 572 & 1~017  & 264  \\ 
10 & 24 & 3 & 70 & 99  & 58 & 326 & 87 & 575 & 986   & 318 \\ 
500 & 17 & 3 & 87 & 107 & 116  & 230 & 86 & 601 & 917  & 806 \\ 
aucun tri   & -  & -  & 100 & 100 & 119 & - & - & 689 & 689  & 1~552 \\\hline
\end{tabular}
\end{scriptsize}
\caption{Temps du tri par bloc, le temps d'indexation et la taille des index pour $k=1$ (temps en secondes et taille en Mo), seules trois dimensions
par table sont utilisées.
}
\label{tab:SortBlockTime}
\end{table*}

\begin{figure*}[!t]
\centering
\subfloat[DBGEN]{\includegraphics[width=0.45\textwidth]{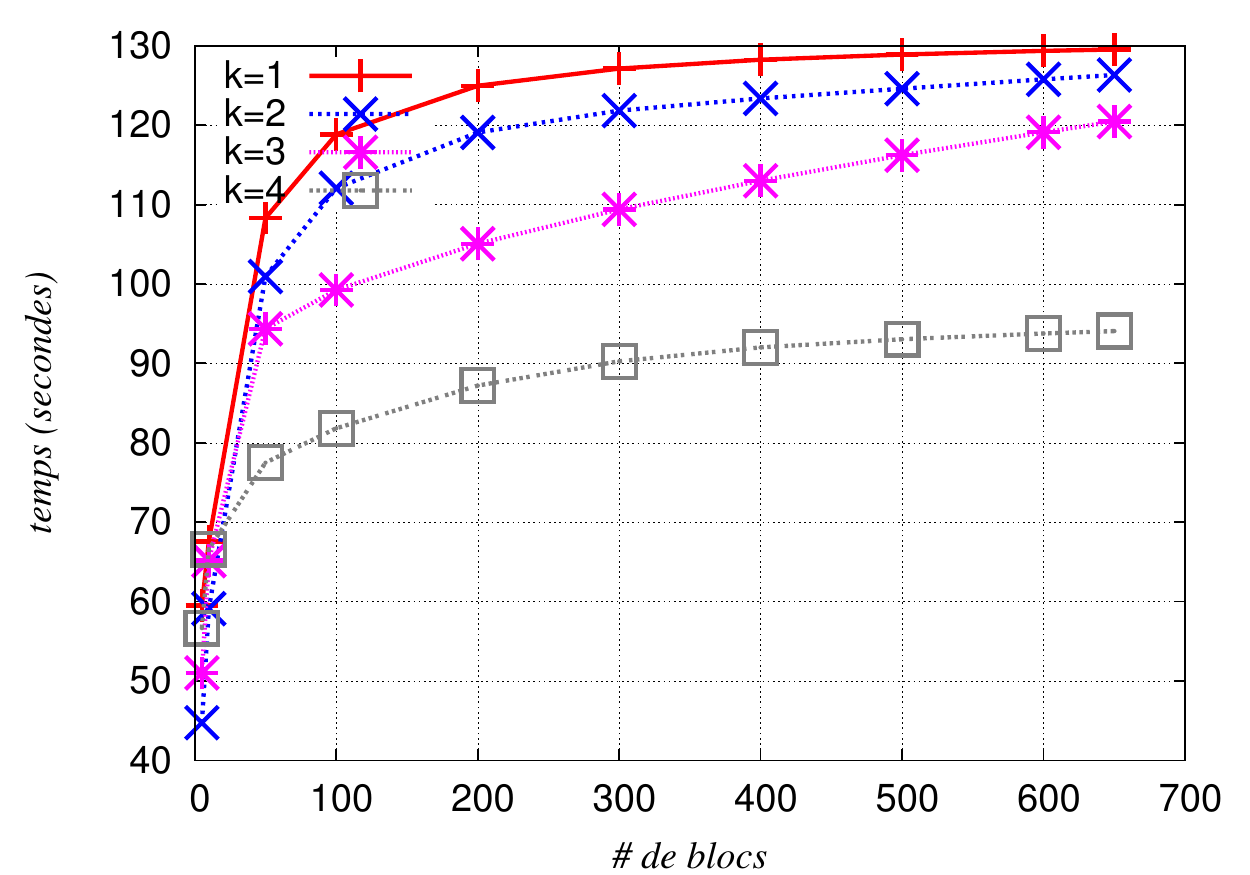}\label{fig:BlocDbgenD3Time}}
\subfloat[Netflix]{\includegraphics[width=0.45\textwidth]{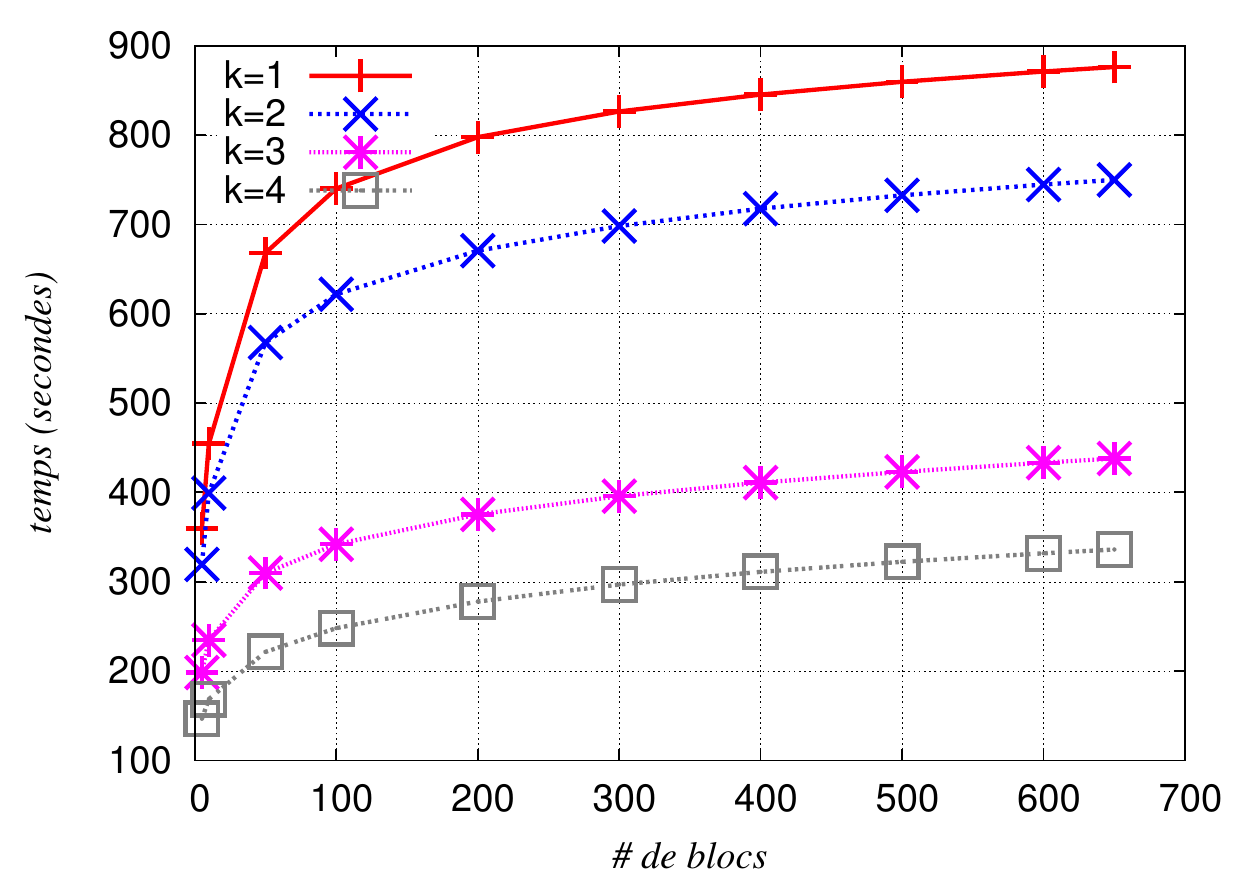}\label{fig:BlocNetflixD3Time}}   
\caption{Temps de construction des index bitmaps en fonction du nombre de blocs utilisés pour le tri, excluant le temps requis pour le tri. } \label{fig:BlockTime}
\end{figure*}

\begin{figure*}[!t]
\centering
\subfloat[DBGEN]{\includegraphics[width=0.45\textwidth]{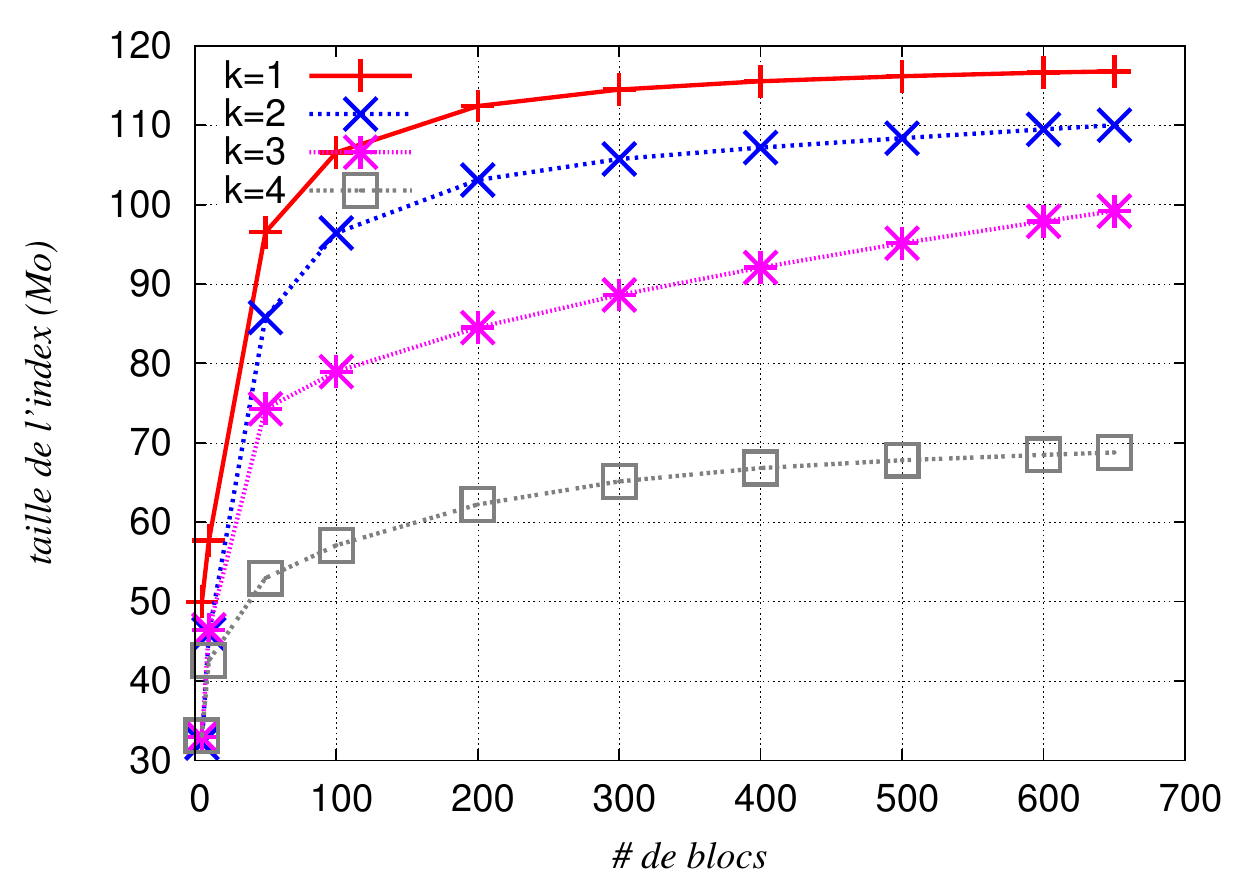}\label{fig:BlocDbgenD3Size}} 
\subfloat[Netflix]{\includegraphics[width=0.45\textwidth]{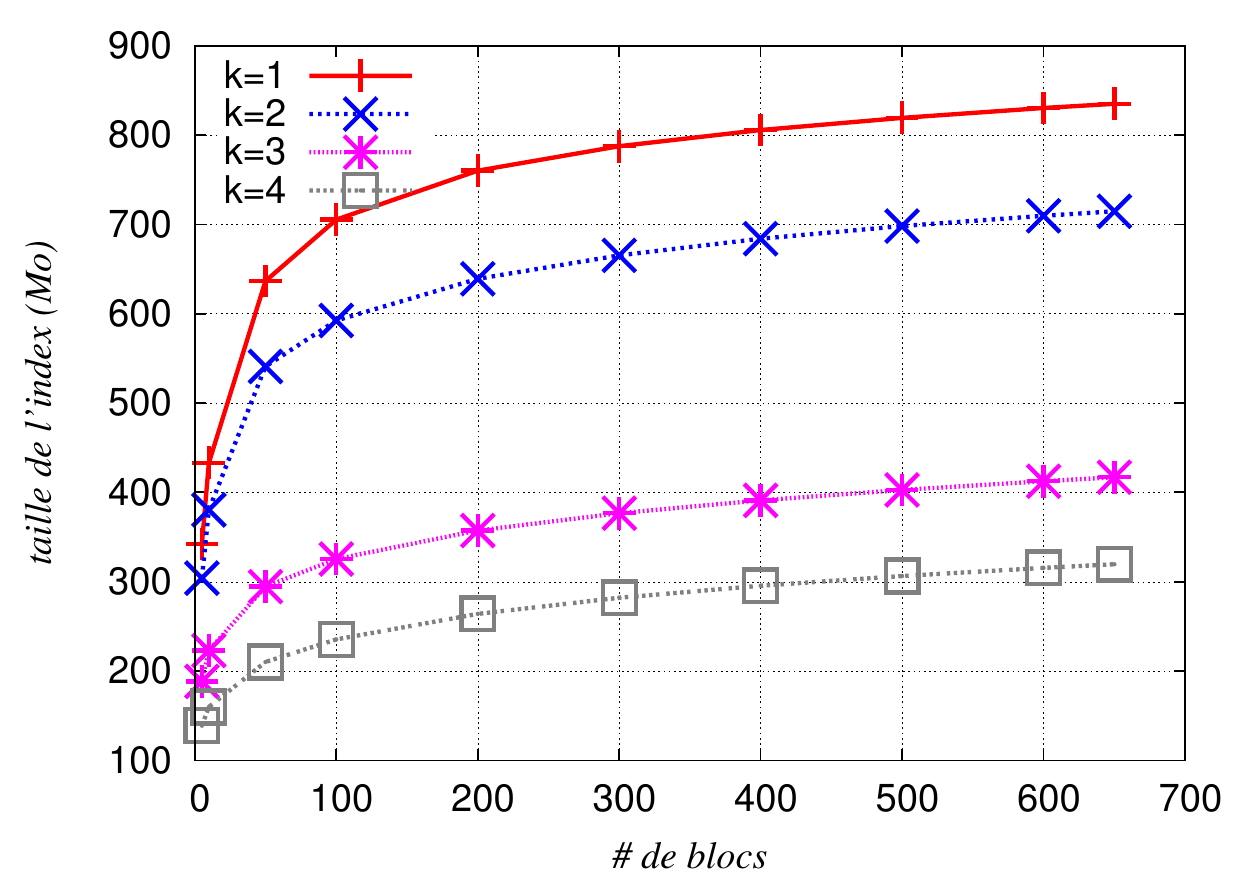}\label{fig:BlocNetflixD3Size}}
\caption{Taille des index en fonction du nombre de blocs utilisés pour le tri.} \label{fig:BlockSize}
\end{figure*}

Les Figures~\ref{fig:BlockTime}~et~\ref{fig:BlockSize} montrent la variation du temps de création
des index et leur taille en fonction du nombre de blocs utilisés pour trier la table de faits. Le temps
de création des index (pris à chaud, donc sans le temps de calcul de l'histogramme) et l'espace qu'ils
occupent augmentent proportionnellement  
 au nombre de blocs.
 Le traitement par bloc, sans fusion des blocs, empêche certains regroupement de
 valeurs d'attributs. Plus il y a de blocs,  moins il y de de regroupements possibles
et par conséquent moins de compression. 
 \cut{
 \kamel{Pour moi la phrase est compréhensible mais tu peux bien sûr modifier la phrase et le paragraphe.}
 \daniel{Vois  le changement que j'ai fait, mais ce n'est pas tellement que ta fin de paragraphe n'était
 pas compréhensible (le français était clair). Je trouve que ça pourrait mieux couler par contre.
 Enfin. Vois ce que j'ai mis et voyons si on peut arriver à quelque chose qui nous plaît à tous les deux.}
  En effet, lorsqu'on augmente le
nombre de blocs, on a moins de cas où des faits contigus qui ont les mêmes valeurs d'attributs pour chaque
dimension. Ce qui revient à dire qu'on a moins de possibilités de regroupement des bitmaps
et par conséquent moins de compression.  
 } Cela donne alors lieu à des index bitmaps plus volumineux. L'effet est considérable~: le fait de
 trier Netflix en 5~blocs peut doubler la taille de l'index (voir Tableau~\ref{tab:SortBlockTime})
 par rapport à un tri complet. 
Le temps de construction est plus élevé quand le nombre de blocs augmente, car on se retrouve avec plus de données à écrire sur disque.

Par ailleurs, pour un nombre de blocs donnée, la taille des index et le temps de construction varient
inversement par rapport à la valeur de $k$ utilisée pour l'encodage des bitmaps. L'effet de $k$ sur la taille
et le temps vient du fait que lorsque $k$ a une petite valeur, le nombre de bitmaps est plus élevé. 
Un index
plus petit est construit plus rapidement.
Cependant, parce que des bitmaps moins nombreux, mais plus sont moins compressibles, la taille de l'index n'est
 pas proportionnelle au nombre de bitmaps. En effet, alors que le nombre de bitmaps décroît exponentiellement
avec $k$, la taille de l'index n'est réduite que d'un facteur de 2 à 4 lorsque $k$ passe de 1 à 4.

%
Nous présentons le temps d'interrogation des données via les index bitmaps en fonction du nombre de blocs utilisés
pour le tri. 
Pour cela, nous avons sélectionné pour chaque jeu de données 12 requêtes d'égalité ayant des
sélectivités différentes. Chaque requête est définie sur une seule dimension dont 
la cardinalité est élevée. 



\begin{figure*}[!t]  
  \subfloat[DBGEN (100~000 exécutions)]{\includegraphics[width=0.45\textwidth]{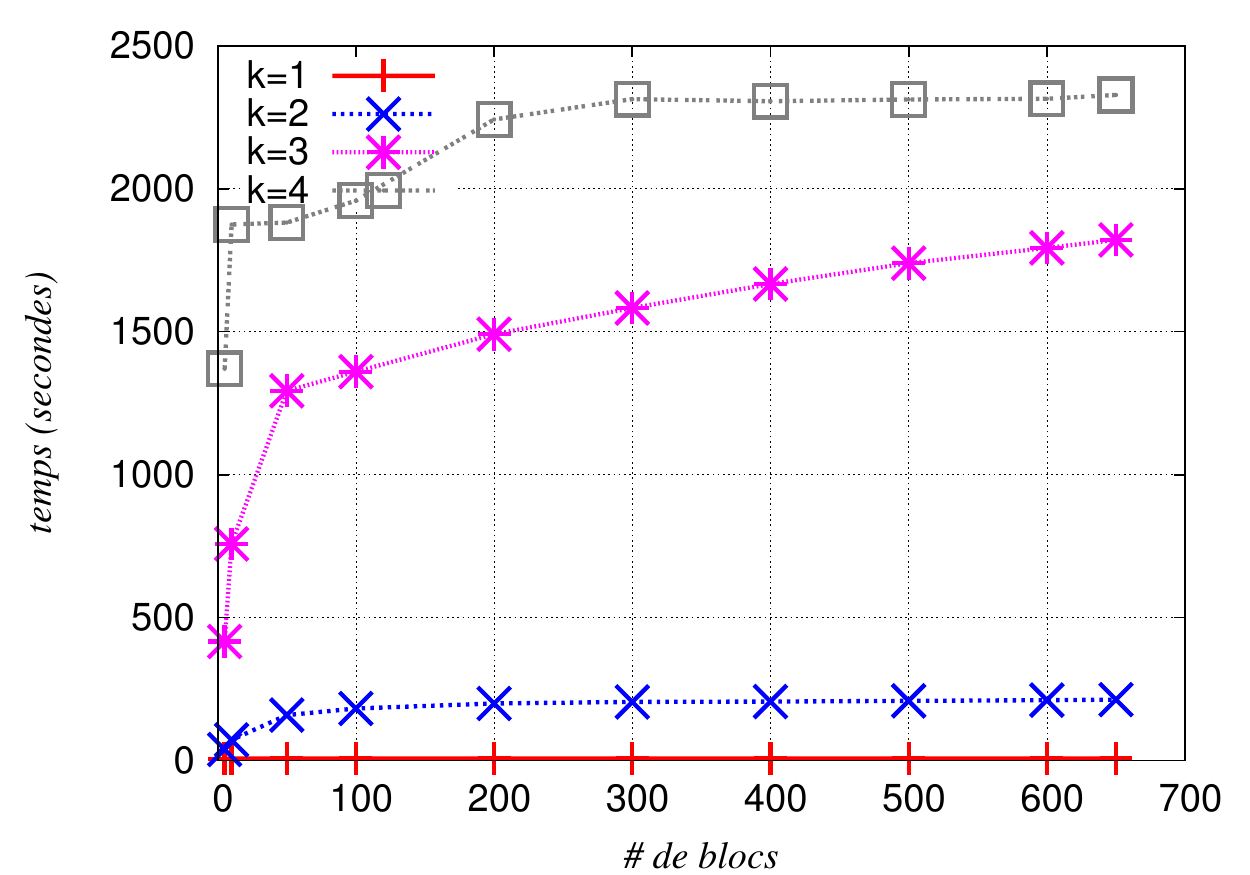}\label{fig:QBDbgen}} 
  \subfloat[Netflix (30 exécutions)]{\includegraphics[width=0.45\textwidth]{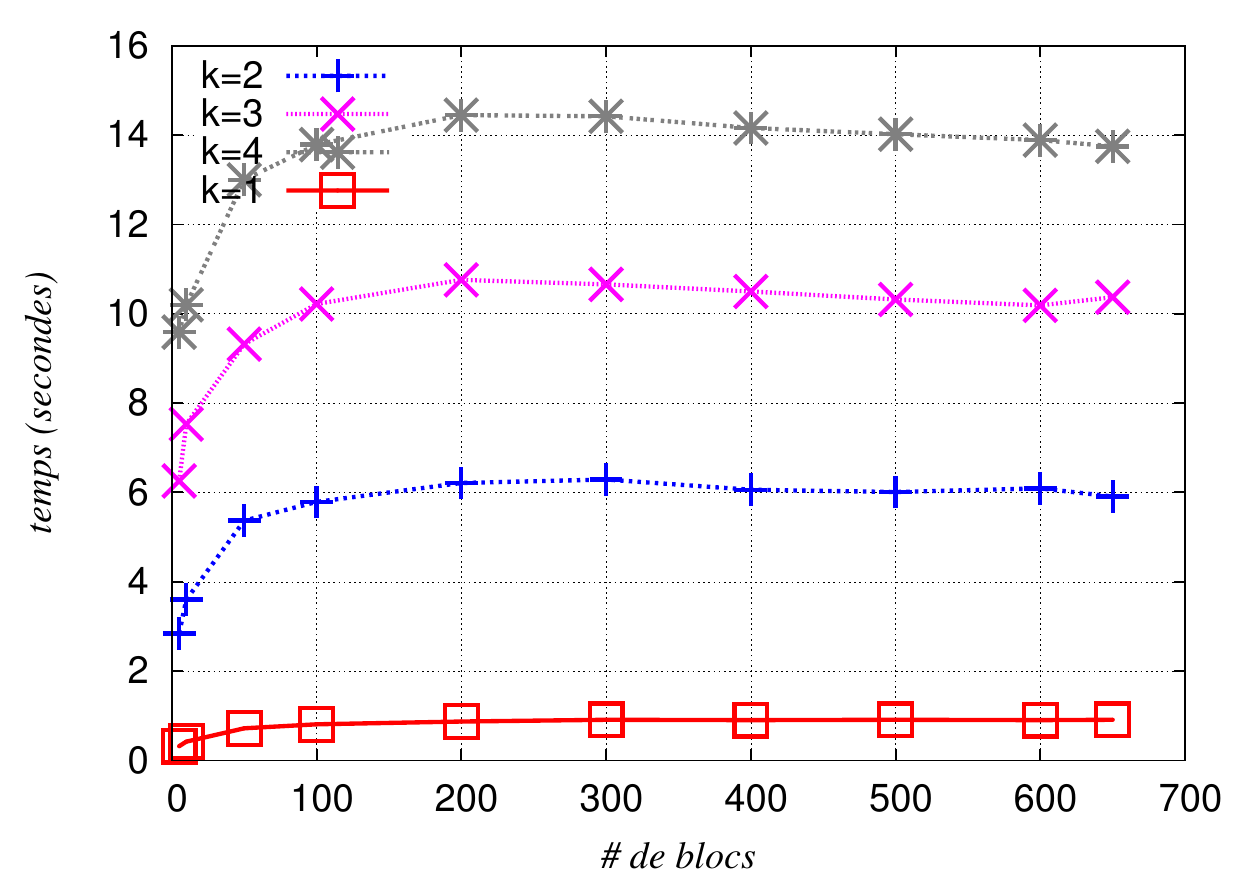}\label{fig:QBNetflix}}
\caption{Temps d'interrogation des données via les index bitmaps en fonction du nombre de blocs utilisés pour trier les données.
} 
\label{fig:QB}
\end{figure*}

Nous constatons à partir de la Figure~\ref{fig:QB} que le temps d'interrogation augmente en fonction
du nombre de blocs, pour devenir presque constant pour les plus grandes valeurs du nombre de blocs. De plus, pour un nombre de blocs donné, le temps d'interrogation est inversement proportionnel à la valeur de $k$ utilisée pour l'encodage des bitmaps. En effet, lorsque la valeur de $k$ est grande, on consulte plus de bitmaps ($k$ bitmaps) pour chaque dimension. 
Le fait de passer de $k=1$ à $k=2$ 
multiplie le temps de traitement par un facteur de 6 ou plus, même si le nombre de bitmaps consulté
n'est que doublé.
 Il y a donc un compromis à faire entre le temps de construction et l'espace occupé par l'index d'une part, et la performance de l'index d'autre part.

\danielcut{
Par ailleurs, les Figure~\ref{fig:QS} montrent que le temps d'interrogation augmente en fonction de la sélectivité.
 Plus celle-ci est grande, plus on a de données à chercher et donc de bits mis à 1 à retrouver dans les bitmaps
  des index \kamel{un peu simpliste comme explication mais je trouverai mieux}.
}

\cut{ from Kamel
# of runs (second parameter) by default is 30,
nohup ./querybyblocks.sh /Volumes/eric/netflix/netflix.r.csv.d023
"2:2005-01-19 2:2005-04-06 2:2005-06-01 2:2005-06-27 2:2005-05-31
2:2005-08-22 2:2005-06-06 2:2005-09-26 2:2005-01-25 2:2005-07-11
2:2005-06-07 2:2005-01-10 "&
nohup ./querybyblocks.sh ../data/usincome/census-income.data.d5240
"0:0 0:500 0:600 0:700 0:800 0:1000 0:425 0:900 0:550 0:1200 0:1100
0:650" 10000&
nohup ./querybyblocks.sh ../data/dbgen/lineitem.txt.r.d136 "0:250977
0:106436  0:263656  0:9933  0:343409  0:321459  0:133004  0:218786
0:357532  0:194699  0:4916  0:236346" 100000&
}

\danielcut{
\begin{figure*}[!t]
\centering
  \subfloat[DBGEN]{\includegraphics[width=0.50\textwidth]{query_sel_dbgen}\label{fig:QSDbgen}}
  \subfloat[Netflix]{\includegraphics[width=0.50\textwidth]{query_sel_netflix}\label{fig:QSNetflix}}
\caption{Temps d'interrogation des données via les index bitmaps en fonction de la sélectivité des requêtes.
 } \label{fig:QS}
\end{figure*}
}

Au total, il semble que le remplacement du tri complet par un tri par bloc ne soit pas une
bonne stratégie pour des tables de faits ayant moins que des centaines de millions de faits.
En effet, le Tableau~\ref{tab:SortBlockTime} montre que si le tri par bloc peut être 20\,\% plus rapide, 
 il peut aussi générer des index plus volumineux qui prennent plus de temps de construction.
 En fait,   le tri par bloc peut même causer une augmentation
du temps total d'indexation. Les performances des index résultant d'un tri par bloc sont moindres selon
la Figure~\ref{fig:QB}.

\section{Conclusion et perspectives}

Nous avons utilisé le tri de la table de faits pour réduire la taille des index bitmaps. 
À l'aide du tri, nous pouvons réduire par un facteur de deux  la taille d'un index, tout en doublant parfois la performance
de l'index. Dans les cas où il y a un grand nombre de colonnes, ces gains se concentrent surtout sur 
les premières colonnes. 

Les techniques de tri et d'allocation des bitmaps que nous avons considérées n'exploitent pas
l'histogramme des données. Nous travaillons à améliorer l'effet du tri en tenant compte
de la distribution de la fréquence des valeurs d'attributs. Selon nos tests préliminaires, certains gains supplémentaires 
sont possibles.

Notre table de faits la plus grande comportait 100~millions de lignes. Pour les tables
beaucoup plus grandes, ayant des milliards ou des billions de faits, un tri complet de
la table peut s'avérer beaucoup plus coûteux que le temps de création de l'index. En effet,
l'indexation se fait en temps linéaire par rapport au nombre de fait ($O(n)$) alors
que le tri requière un temps $O(n\log n)$. Les
solutions alternatives comme le tri par bloc sans fusion ou le simple regroupement 
des valeurs similaires pourraient s'avérer alors plus
intéressantes. Nous
prévoyons de traiter ces cas dans nos futurs travaux.

\cut{
\owen{sometimes gc sorting of the index, AFTER permuting columns
and inverting a random subset of columns, works well.  Why? }
\daniel{Care to elaborate Owen?}
\owen{Maybe we should just keep this totally secret till the next paper.
I don't know whether this is just an effect on my synthetic data,
or whether it holds on real data.  Essentially, this is the basis
of my emailed belief that interleaving the bitmaps for dimensions will
be a big winner. By giving up doing a good job on the first dimension
or two, you end up getting a small improvement on all dimensions,
and it seems to work out better.  (I am assuming that the reordering
of bitmaps is significant, and the sorting effect of inverting some
is negligible.  But that is just a guess.)
Visualizing this with the spiky per-bitmap diagrams would be appropriate.,
as well as getting dimensionaldirty outputs.}
}





\bibliographystyle{fr-abbrv} 
\bibliography{../bib/lemur}

\begin{thebibliography}{10}

\bibitem{874730}
G.~Antoshenkov.
\newblock Byte-aligned bitmap compression.
\newblock Dans {\em DCC'95}, page 476, 1995.

\bibitem{xldb07---fr}
J.~Becla et K.-T. Lim.
\newblock Report from the 1st workshop on extremely large databases.
\newblock En ligne~:
  \url{http://www-conf.slac.stanford.edu/xldb07/xldb07_report.pdf}, 2007.
\newblock Dernier acc\`es le 22 avril 2008.

\bibitem{pinarunpublished---fr}
G.~Canahuate, H.~Ferhatosmanoglu, et A.~Pinar.
\newblock Improving bitmap index compression by data reorganization.
\newblock En ligne~: \url{http://hpcrd.lbl.gov/~apinar/papers/TKDE06.pdf},
  2006.
\newblock Dernier acc\`es le 12 f\'evrier 2008.

\bibitem{chan1998bid}
C.~Y. Chan et Y.~E. Ioannidis.
\newblock Bitmap index design and evaluation.
\newblock Dans {\em SIGMOD'98}, pages 355--366, 1998.

\bibitem{chan1999ebe}
C.~Y. Chan et Y.~E. Ioannidis.
\newblock An efficient bitmap encoding scheme for selection queries.
\newblock Dans {\em SIGMOD'99}, pages 215--226, 1999.

\bibitem{Christofides1976}
N.~Christofides.
\newblock Worst-case analysis of a new heuristic for the travelling salesman
  problem.
\newblock Technical Report 388, Graduate School of Industrial Administration,
  Carnegie Mellon University, 1976.

\bibitem{KDDRepository---fr}
S.~Hettich et S.~D. Bay.
\newblock The {UCI} {KDD} archive.
\newblock En ligne~: \url{http://kdd.ics.uci.edu}, 2000.
\newblock Dernier acc\`es le 21 d\'ecembre 2007.

\bibitem{qdbm---fr}
M.~Hirabayashi.
\newblock {QDBM}: Quick database manager.
\newblock En ligne~: \url{http://qdbm.sourceforge.net/}, 2006.
\newblock Dernier acc\`es le 22 f\'evrier 2008.

\bibitem{johnson2004clb}
D.~Johnson, S.~Krishnan, J.~Chhugani, S.~Kumar, et S.~Venkatasubramanian.
\newblock Compressing large boolean matrices using reordering techniques.
\newblock Dans {\em VLDB'04}, pages 13--23, 2004.

\bibitem{jurgens2001tbi}
M.~Jurgens et H.~J. Lenz.
\newblock Tree based indexes versus bitmap indexes: A performance study.
\newblock {\em International Journal of Cooperative Information Systems},
  10(3):355--376, 2001.

\bibitem{kimball1996dwt}
R.~Kimball.
\newblock {\em The data warehouse toolkit: practical techniques for building
  dimensional data warehouses}.
\newblock John Wiley \& Sons, Inc. New York, NY, USA, 1996.

\bibitem{354819}
N.~Koudas.
\newblock Space efficient bitmap indexing.
\newblock Dans {\em CIKM '00}, pages 194--201, 2000.

\bibitem{DBLPXML---fr}
M.~Ley.
\newblock \url{http://dblp.uni-trier.de/xml/}, 2008.
\newblock Dernier acc\`es en mars 2008.

\bibitem{133210}
A.~Moffat et J.~Zobel.
\newblock Parameterised compression for sparse bitmaps.
\newblock Dans {\em SIGIR'92}, pages 274--285, 1992.

\bibitem{monashbigtera2007}
C.~Monash.
\newblock Datallegro heads for the high end.
\newblock
  \url{http://www.dbms2.com/2007/07/25/datallegro-heads-for-the-high-end/},
  2007.

\bibitem{netflixprize---fr}
{Netflix, Inc.}
\newblock Nexflix prize.
\newblock \url{http://www.netflixprize.com}, 2007.
\newblock Dernier acc\`es le 4 d\'ecembre 2007.

\bibitem{658338}
P.~E. O'Neil.
\newblock Model 204 architecture and performance.
\newblock Dans {\em 2nd International Workshop on High Performance Transaction
  Systems}, pages 40--59, 1989.

\bibitem{pinar05}
A.~Pinar, T.~Tao, et H.~Ferhatosmanoglu.
\newblock Compressing bitmap indices by data reorganization.
\newblock Dans {\em ICDE'05}, pages 310--321, 2005.

\bibitem{1155030}
D.~Rotem, K.~Stockinger, et K.~Wu.
\newblock Minimizing {I/O} costs of multi-dimensional queries with bitmap
  indices.
\newblock Dans {\em SSDBM '06}, pages 33--44, 2006.

\bibitem{oraclevivekbitmap---fr}
V.~Sharma.
\newblock Bitmap index vs. b-tree index: Which and when?
\newblock En ligne~:
  \url{http://www.oracle.com/technology/pub/articles/sharma_indexes.html}, mars
  2005.
\newblock Dernier acc\`es le 22 avril 2008.

\bibitem{stockinger2004esb}
K.~Stockinger, K.~Wu, et A.~Shoshani.
\newblock Evaluation strategies for bitmap indices with binning.
\newblock Dans {\em {DEXA} '04}, 2004.

\bibitem{DBGEN---fr}
{TPC}.
\newblock {DBGEN} 2.4.0.
\newblock En ligne~: \url{http://www.tpc.org/tpch/}, 2006.
\newblock Dernier acc\`es le 4 d\'ecembre 2007.

\bibitem{671192}
R.~Weber, H.-J. Schek, et S.~Blott.
\newblock A quantitative analysis and performance study for similarity-search
  methods in high-dimensional spaces.
\newblock Dans {\em VLDB '98}, pages 194--205, 1998.

\bibitem{502689}
K.~Wu, E.~J. Otoo, et A.~Shoshani.
\newblock A performance comparison of bitmap indexes.
\newblock Dans {\em CIKM '01}, pages 559--561, 2001.

\bibitem{wu2006obi}
K.~Wu, E.~J. Otoo, et A.~Shoshani.
\newblock Optimizing bitmap indices with efficient compression.
\newblock {\em ACM Transactions on Database Systems}, 31(1):1--38, 2006.

\bibitem{yiannis2007ctf}
J.~Yiannis et J.~Zobel.
\newblock Compression techniques for fast external sorting.
\newblock {\em The VLDB Journal}, 16(2):269--291, 2007.

\end{thebibliography}
\end{document}